\documentclass[letter]{aa}     

\usepackage{graphicx}
\usepackage{txfonts}
\usepackage{lipsum}
\usepackage{subcaption}         
\usepackage{lscape}             
\usepackage{placeins}           

\usepackage[]{hyperref}

\begin{document}

   \title{Testing a proposed ''planarity'' tool for studying satellite systems}

   \subtitle{On the alleged consistency of Milky Way satellite galaxy planes with $\Lambda$CDM}

   \author{Marcel S. Pawlowski\inst{1}
          \and
          Mariana P. Júlio \inst{1, 2}
          \and
          Kosuke Jamie Kanehisa \inst{1, 2}        
          \and
          Oliver Müller \inst{3}
          }

   \institute{Leibniz-Institute for Astrophysics Potsdam (AIP), An der Sternwarte 16, 14482 Potsdam, Germany\\
              \email{mpawlowski@aip.de}
         \and
             Institut für Physik und Astronomie, Universität Potsdam, Karl-Liebknecht-Straße 24/25, D-14476 Potsdam, Germany  
        \and
             Institute of Physics, Laboratory of Astrophysics, Ecole Polytechnique Fédérale de Lausanne (EPFL), 1290 Sauverny, Switzerland
             \\ }

   \date{Received December XX, 2024}

 
  \abstract
   {The existence of planes of satellite galaxies has been identified as a long-standing challenge to $\Lambda$CDM cosmology, due to the rarity of satellite systems in cosmological simulations that are as extremely flattened and as strongly kinematically correlated as observed structures.}
   {Here we investigate a recently proposed new metric to measure the overall degree of ''planarity'' of a satellite system, which was used to claim consistency between the Milky Way satellite plane and $\Lambda$CDM.} 
   {We study the behavior of the ''planarity'' metric under several features of anisotropy present in $\Lambda$CDM satellite systems but unrelated to satellite planes. Specifically, we consider the impact of oblate or prolate distributions, the number of satellites, clustering of satellites, and radial and asymmetric distributions ('lopsidedness'). We also investigate whether the metric is independent of the orientation of the studied satellite system.}
   {We find that all of these features of anisotropy result in the metric inferring an increased degree of ''planarity'', despite none of them having any direct relation to satellite planes. The metric is also highly sensitive to the orientation of the studied system (or chosen coordinate system): there is almost no correlation between the metric's reported degrees of ''planarity'' for identical random systems rotated by $90^\circ$.}
   {Our results demonstrate that the new proposed metric is unsuitable to measure overall ''planarity'' in satellite systems. Consequently, no consistency of the observed Milky Way satellite plane with $\Lambda$CDM can be inferred using this metric.}

   \keywords{galaxy: formation -- galaxies: dwarf -- galaxies: evolution -- galaxies: structure -- methods: data analysis
               }

   \maketitle

\section{Introduction}

Observational evidence for the presence of planes of satellite galaxies that likely co-orbit has been demonstrated for numerous systems (see \citealt{2018MPLA...3330004P} for a review). Well studied cases include the Milky Way \citep{2005A&A...431..517K, 2012MNRAS.423.1109P, 2024A&A...681A..73T}, M31 \citep{2013Natur.493...62I, 2020ApJ...901...43S}, Centaurus A \citep{2015ApJ...802L..25T,2018Sci...359..534M, 2023MNRAS.519.6184K}, and NGC4490 \citep{2024MNRAS.528.2805K}. 
For these observed systems, analogs with similar degrees of spatial flattening and kinematic coherence are rare in $\Lambda$CDM simulations \citep{2014ApJ...784L...6I, 2014ApJ...789L..24P, 2018MNRAS.478.5533F, 2019ApJ...875..105P, 2021A&A...645L...5M, 2021ApJ...923...42P, 2021MNRAS.504.1379S, 2024A&A...688A.153P, 2024ApJ...976..253S}. 

Other observed host galaxies also show some signs of possible planarity or kinematic coherence in their associated satellites, but the degree of tension with cosmological expectations is less well established (e.g. \citealt{2013AJ....146..126C, 2021ApJ...917L..18P, 2021A&A...652A..48M, 2024A&A...683A.250M, 2024ApJ...966..188M, 2024arXiv240503769M}).

Spatial flattening is commonly measured as the major-to-minor axis ratio in 2D, or the absolute root-mean-square plane height in 3D. The kinematic coherence is either measured as the dispersion of orbital poles if 3D velocities are known from proper motions, or as 2D line-of-sight velocity trends for more distant systems.
In the presence of well established methods used widely by many different teams, introducing new metrics to measure satellite planes \citep{2019MNRAS.488.1166S, 2022arXiv220805496F, 2024ApJ...976..253S}
can hinder comparability. When a novel tool is also applied only to a new simulation, it prevents judgment on whether an apparent consistency between observation and simulation is due to the latter being more successful in reproducing the observed system than previous simulations, or rather due to shortcomings of the new metric (which often affect these proposed new analysis tools, see e.g. \citealt{2014MNRAS.442.2362P, 2015ApJ...815...19P, 2017AN....338..854P,2020MNRAS.491.3042P}).

We therefore need to ensure that the tools we apply are suitable and have been demonstrated to reliably measure what we want them to measure. A good tool needs to have both a high sensitivity and a high specificity. The former implies that the tool can accurately diagnose a property present in the data (such as the presence of planes in the distribution of satellite galaxies), while the latter requires that the tool does not return false positive diagnoses in the absence of the condition being tested for (such as reporting high degrees of planarity for systems without intrinsic planes).

\citet{2024arXiv241117813U} have recently proposed yet another new tool, aimed at measuring the overall ''planarity'' present in a system of satellite galaxies. 
The metric intends to measure -- with a single value -- the overall degree of planarity in a distribution. Their new tool does return high degrees of ''planarity'' not only for the distribution of Milky Way satellite galaxies, but also for a range of satellite systems extracted from the NewHorizon cosmological simulation \citep{2021A&A...651A.109D}. \citet{2024arXiv241117813U} interpret this as demonstrating consistency between $\Lambda$CDM expectations and the observed system and its satellite plane.

By generating one mock satellite system in which satellite sub-samples are confined to three planes, \citet{2024arXiv241117813U} demonstrate the tool's sensitivity to the presence of planar distributions. However, reverse tests investigating the specificity have not been presented. It is thus unclear whether the proposed tool is a suitable metric to reliably measure ''planarity'', or might instead be affected by other influences, such as by different types of deviations of the satellite systems from isotropy. 

In the following, we investigate the metric and its response to several types of phase-space correlations present in satellite galaxy systems, both observed in the Universe and extracted from $\Lambda$CDM simulations. 
We show that the proposed metric lacks specificity, since it is sensitive to other anisotropies that are independent of the presence of planar arrangements, a result contrary to its intended purpose. We also demonstrate that the degree of ''planarity'' it returns is affected by the orientation of its coordinate system relative to the studied satellite distribution.
We refrain from commenting on the application of the metric on velocity vectors, because the following investigations of positions alone already disqualify it from further use. However, we note that the procedure employed by \citet{2024arXiv241117813U} to sample from the measurement uncertainties of observed satellite galaxy positions and velocities, namely sampling them in 6D Cartesian coordinates independently, ignores the presence of strong correlations between these and results in nonphysical satellite phase-space positions (see Appendix \ref{appendix}).

We note that \citet{2024arXiv241117813U} require each host to contain more than 30 satellites with a stellar masses $> 10^5 M_\odot$, while considering hosts with a stellar mass $> 10^{10} M_\odot$. Given that for the Milky Way we only know of 15 satellites that exceed this stellar mass \citep{2024arXiv241107424P}, it appears plausible that many simulated hosts might be more massive than the Milky Way. Yet, since no information on the host mass distributions or the number of satellites per host is provided in \citet{2024arXiv241117813U}, we can not make definitive statements on this issue.

\section{Investigating the ''planarity'' metric}

The metric proposed by \citet{2024arXiv241117813U} constructs the cross-products of all possible combinations of satellite galaxy position vectors\footnote{In this regard it is similar to the three- and four-galaxies normal methods of \citet{2013ApJ...766..120C} and \citet{2013MNRAS.435.1928P}, who used these as a discovery tool to identify possible sub-sample satellite planes.}. It thus collects all plane normal vectors defined by any combination of two satellites and the center of the coordinate system, which is chosen as the host galaxy position. 

These normal vectors are expressed in spherical coordinates, and binned in $m$\ bins in azimuth and inclination. The resulting 2D-histogram of normal-vector counts per bin is summarized by calculating the Gini coefficient of all bin values. The same is done for 1000 random mock systems, with positions drawn from an isotropic distribution. For a given satellite system under study, its degree of ''planarity'' is reported as the quantile value of its Gini coefficient relative to the distribution of Gini coefficients of these random systems.
\citet{2024arXiv241117813U} report that both the Milky Way system and most simulated satellite systems return very high quantiles, which they interpret as consistency between the observed satellite plane and $\Lambda$CDM.

The reliance on a spherical coordinate system binned in angles implies a special direction in the analysis: the pole of this coordinate system. The orientation of this direction must not affect the metric's output. After all, the presence of planes in a system needs to be measured independently of the orientation under which the system is studied.

Furthermore, different types of phase-space correlations beyond planes are present in both observed satellite systems and those obtained from cosmological simulations (for a review, see \citealt{2021Galax...9...66P}). A metric to measure planarity therefore needs to demonstrate that it does, in fact, measure planarity, and not, for example, just a general deviation from isotropy. This requires testing whether the metric is affected by other phase-space correlations that are independent of the issue of satellite planes.

Since such tests have not been presented by \citet{2024arXiv241117813U}, we set out to do this with a number of toy model systems. For this purpose, we use the code made publicly available by the authors\footnote{Link provided in their paper (last accessed by us on Dec. 12, 2024): https://emiruz.com/vpos, which forwards to https://github.com/emiruz/planarity/}. 
We note that, while \citet{2024arXiv241117813U} describe that the spherical coordinates are scaled to ensure that all bins have equal area, no such scaling is apparent in the provided code. Since we thus can not be sure what procedure has been applied, we here default to the provided code, but repeat our analysis in Appendix \ref{appendix2} after implementing such a scaling. Our main conclusions apply to either case.

Unless stated otherwise, we follow their fiducial choices: $m = 25$ bins per angular dimension, and 1000 isotropic realizations to obtain the quantile of a given system relative to. We base our comparisons on mock systems with $N_\mathrm{sat} = 40$\ satellites, comparable to the number of Milky Way satellites considered by the original study and their requirement that simulated systems contain $N_\mathrm{sat}>30$\ satellites.

   \begin{figure}[h!]
   \centering
   \includegraphics[width=0.9\hsize]{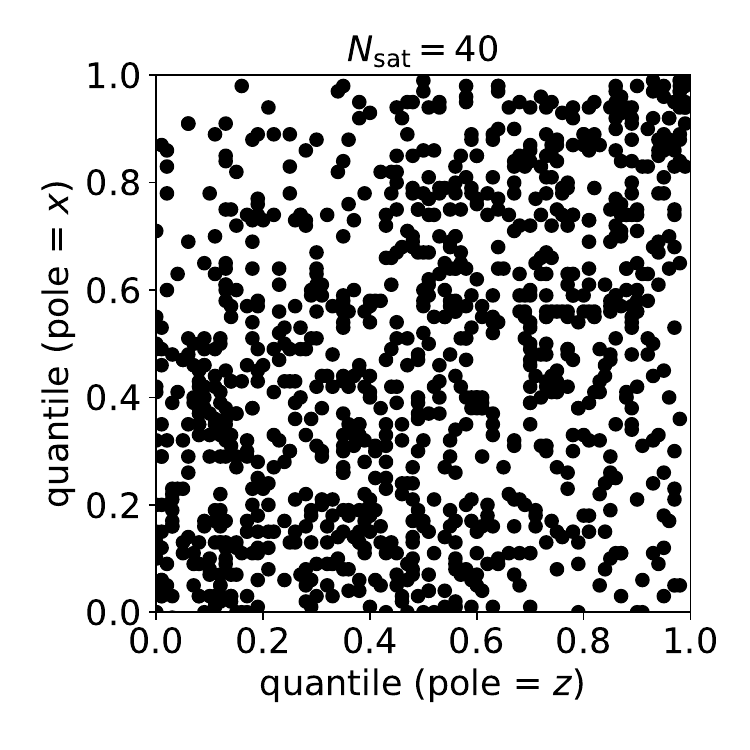}
      \caption{Quantiles for 1000 random isotropic distributions in two orientations rotated by $90^\circ$. No strong correlation is apparent, indicating that the output of the proposed ''planarity'' metric is sensitive to the orientation of the satellite system under study.}
         \label{fig:orientation}
   \end{figure}
   
\subsection{Effect of orientation or coordinate system}

An essential quality of any suitable metric to measure the planarity of a satellite system is that it needs to be independent of the overall orientation of the system under study, or of the chosen coordinate system. The metric relies on a 2-dimensional histogram of the spherical coordinates of pairwise cross-products of vectors. The statistics of the bin counts, this is, how much they deviate from a distribution expected for random systems, is used to quantify the degree of ''planarity''. However, a histogram in two spherical coordinates implies a tighter sampling in azimuth if closer to the poles, and suggests that the orientation of the system might affect the resulting ''planarity'' measure.

We have tested this concern by investigating 1000 random satellite systems, generated as done by \citet{2024arXiv241117813U} by drawing from a homogeneously filled unit sphere, resulting in an isotropic distribution of positions around the origin. For each system we measure the quantile. As expected, random systems result in an overall flat quantile distribution. We then rotate the distributions by $90^\circ$ around the $y$-axis, such that the former $x$-axis lies along the new $z$-axis direction and vice versa. This preserves the mutual distributions of the satellites, only their orientation relative to the $z$-axis defining the orientation of the histogram is different.  We re-run the analysis and again determine the quantiles.

The results of our test are shown in Fig. \ref{fig:orientation}. The quantiles measured in the two orientations vary strongly, with almost no apparent correlation. This is confirmed by tests for both linear and rank correlation: the Pearson correlation coefficient is $r = 0.352$, while the Spearman correlation coefficient is $\rho = 0.351$. A robust metric independent of orientation results in identical quantiles for these rotated systems. The ''planarity'' metric does not. This already shows that the metric is not suitable to study the flattening of a satellite system, but we have uncovered more issues.


   \begin{figure}[h!]
   \centering
   \includegraphics[width=0.9\hsize]{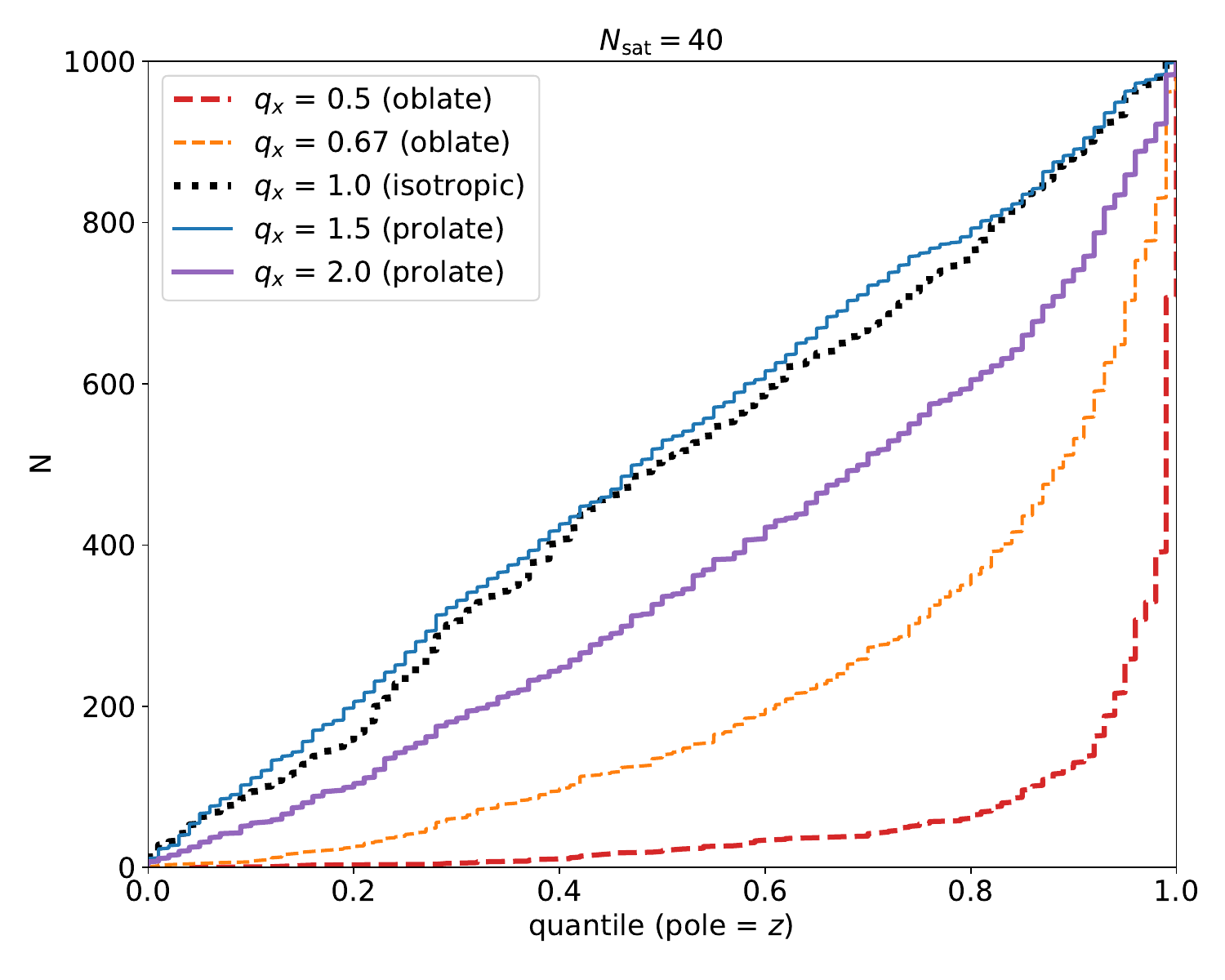}
   \includegraphics[width=0.9\hsize]{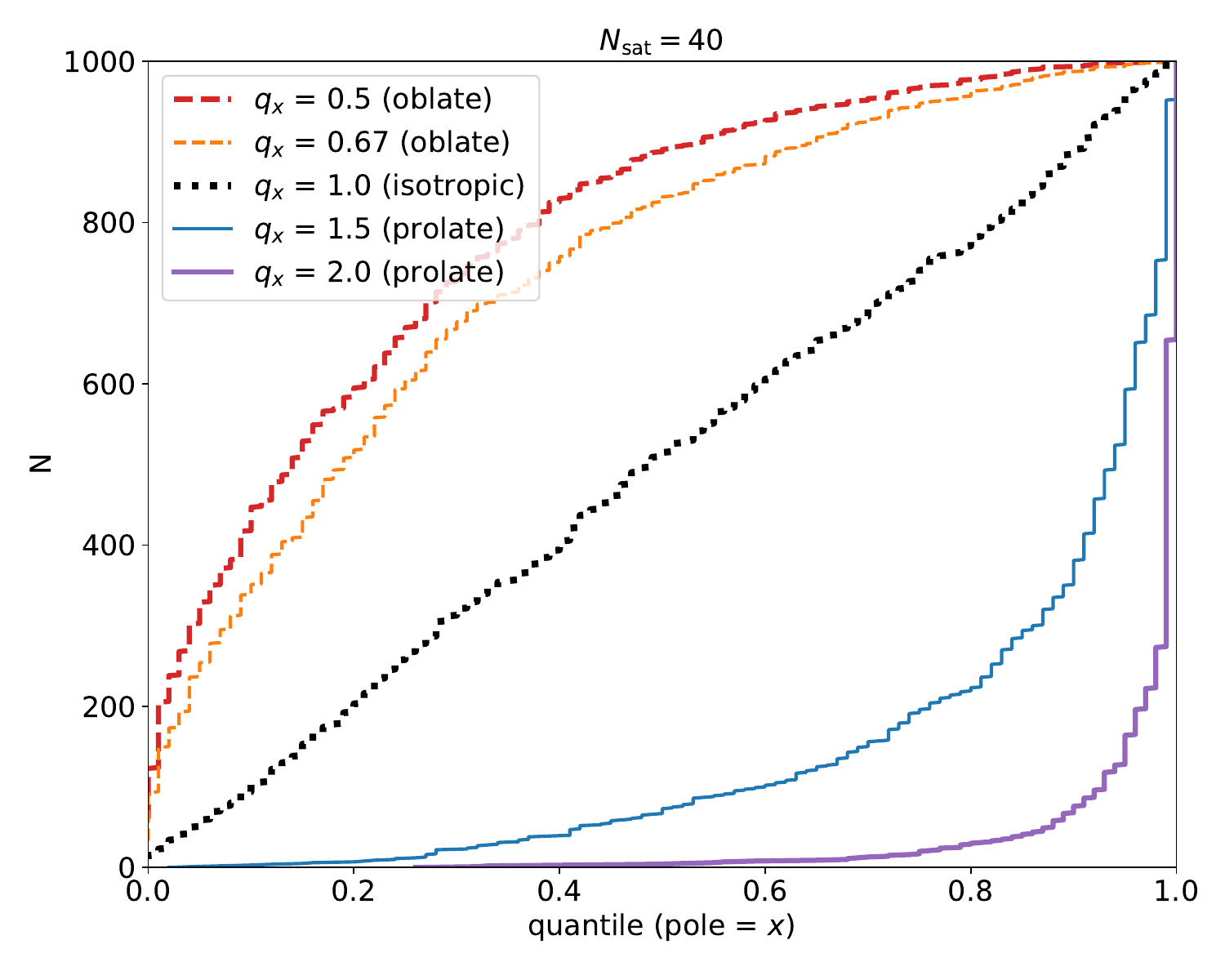}
      \caption{Distribution of quantiles for systems with different degrees of flattening ($q < 1.0$, oblate) or elongation ($q > 1.0$, prolate) along the $x$-axis.
      For the upper panel the metric's pole is oriented along the $z$-, for the lower panel along the $x$-axis.
      }
         \label{fig:shape}
   \end{figure}
   
\subsection{Effect of halo shape}
\label{sect:shape}
   
It is well established that dark matter halos in $\Lambda$CDM, as well as their associated subhalo and thus satellite galaxy systems, are intrinsically tri-axial \citep{2005ApJ...627..647B, 2006MNRAS.367.1781A, 2008MNRAS.385.1511W, 2017MNRAS.467.3226V}. Yet this overall shape does not imply that they contain planes of satellite galaxies. Even though an overall oblate system might be considered to be somewhat plane-like and thus can plausibly be expected to yield a stronger degree of inferred planarity, this is not the result of substantial sub-sample planes as supposedly tested for by the metric. It is thus necessary to test whether the metric is sensitive to the overall shape of the studied systems.

We generate isotropic systems following the fiducial methodology, but re-scale their $x$-axis coordinates by multiplying with a factor $q$. This produces flattened, oblate distributions for $q < 1.0$, and stretched, prolate distributions for $q > 1.0$. We apply the ''planarity'' metric again in two orientations, with its pole along the $z$- and the $x$-axis, respectively. The resulting quantile distributions for 1000 systems generated for each considered $q = [0.5, 0.67, 1.0, 1.5, 2.0]$\ are illustrated in Fig. \ref{fig:shape}. Here, $q = 1.0$\ corresponds to the fiducial, spherical case of \citet{2024arXiv241117813U}.

Our tests show that prolate distributions result in some deviation in the quantile distribution from isotropy, if the direction of stretching and the orientation of the metric's pole are perpendicular. Strongly prolate distributions ($q = 2$) are biased to higher quantile values. Oblate distributions, in contrast, display a more extreme effect and result in a strongly increased number of high-quantile systems. This shows that the proposed 'planarity' metric is highly sensitive to the overall shape of the distribution, even in the absence of underlying embedded satellite planes.

Yet, depending on the orientation, the metric also displays counter-intuitive behavior: If the metric's pole aligns with the $x$-axis (lower panel in Fig \ref{fig:shape}) -- along the direction in which the system is flattened or stretched -- then the metric infers an {\it decreased} degree of planarity for oblate systems. Their quantile distribution becomes heavily skewed to lower values. The metric appears to infer non-planarity if the flattening happens to be oriented perpendicular to the chosen coordinate system's poles. Prolate distributions whose major axis aligns with the pole, in contrast, return very high inferred degrees of planarity, due to a strong bias towards large quantiles. Thus the overall shape of the satellite distribution has a major effect on the quantile returned by the ''planarity'' metric, which can result in high quantile even for systems without intrinsic satellite planes. This feature of the metric will overestimate the frequency of ''planar'' systems in cosmological simulations.

\subsection{Effect of number of satellites}

   \begin{figure*}[h!]
   \resizebox{\hsize}{!}
   {\includegraphics[width=0.32\hsize]{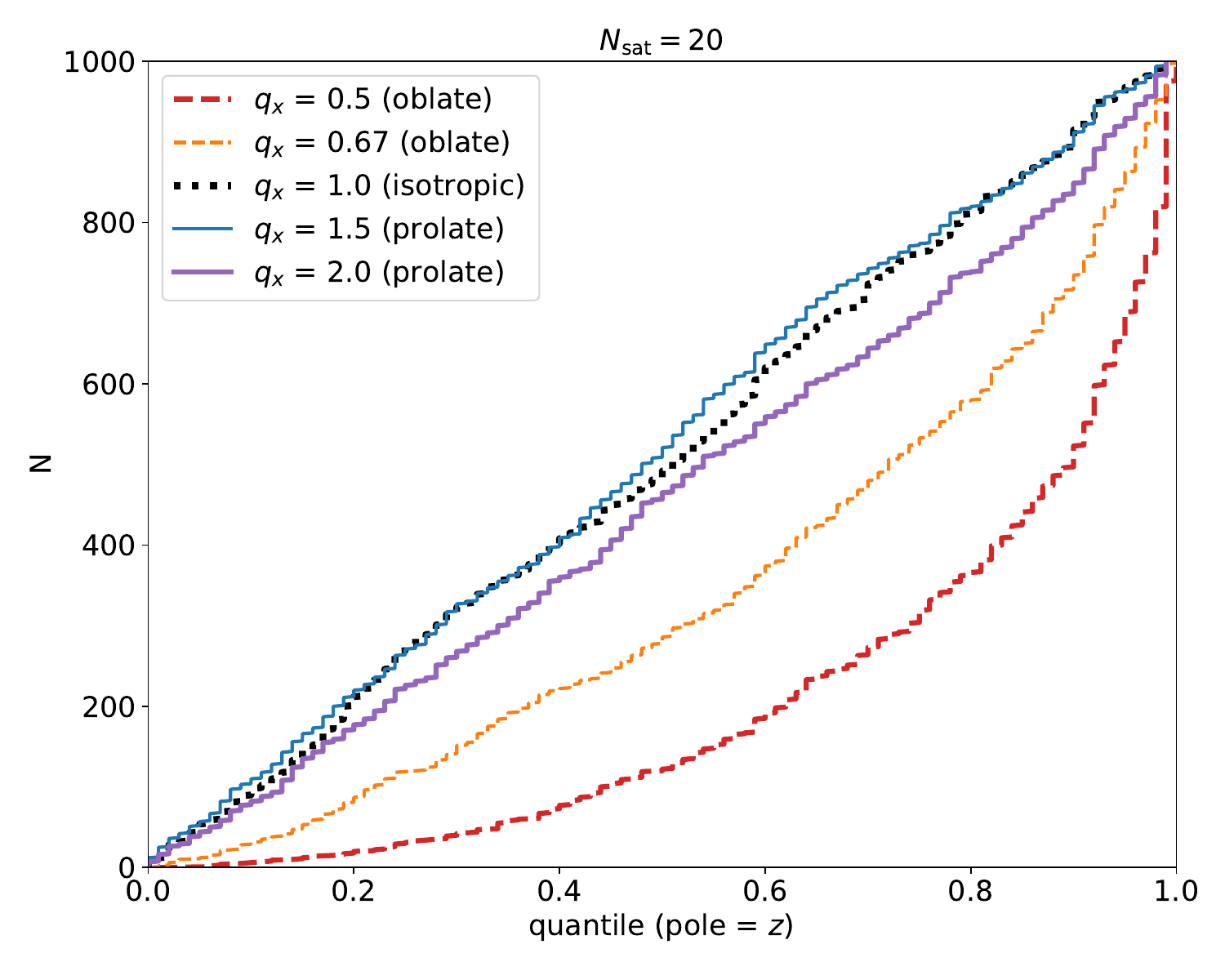}
   \includegraphics[width=0.32\hsize]{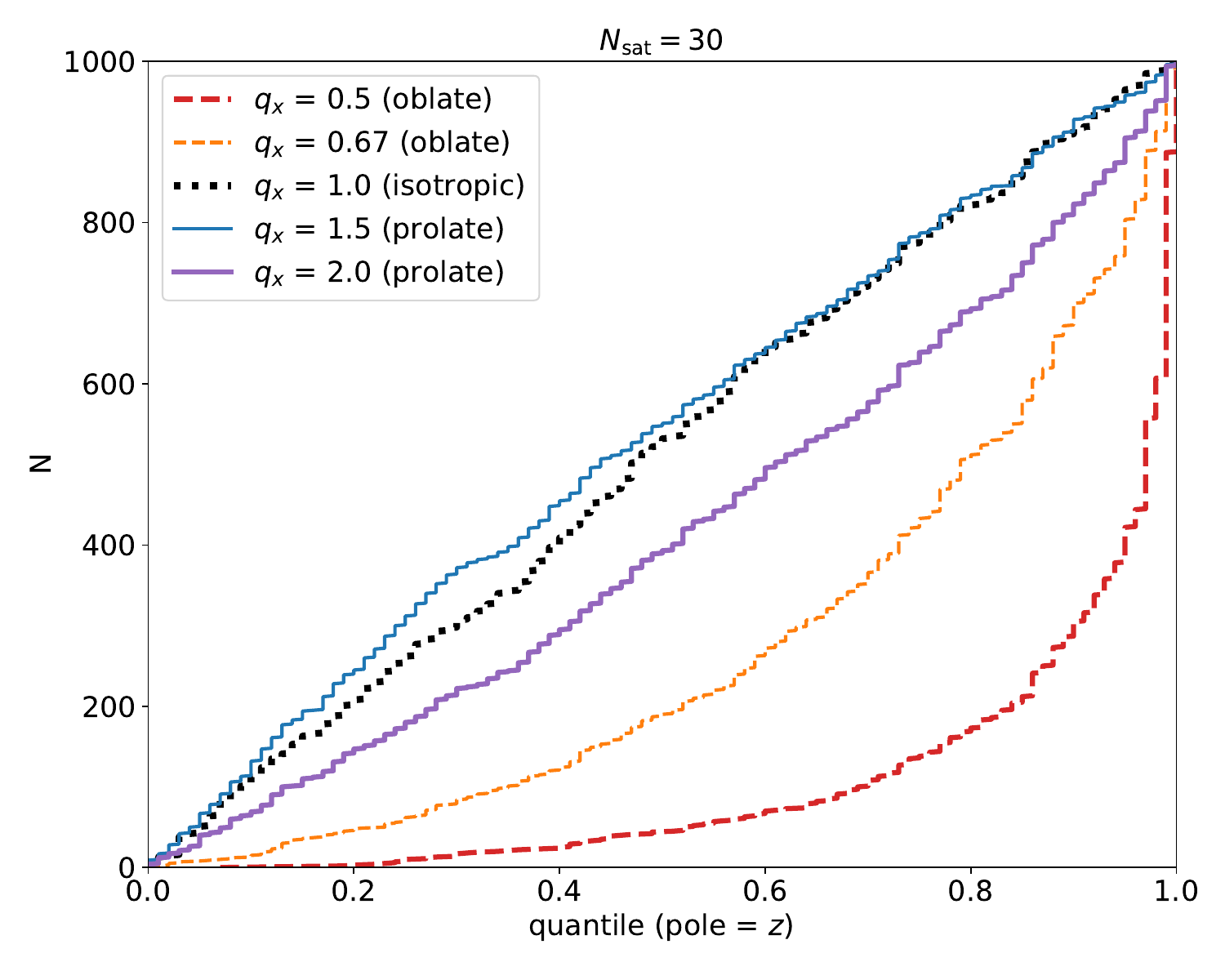}
   \includegraphics[width=0.32\hsize]{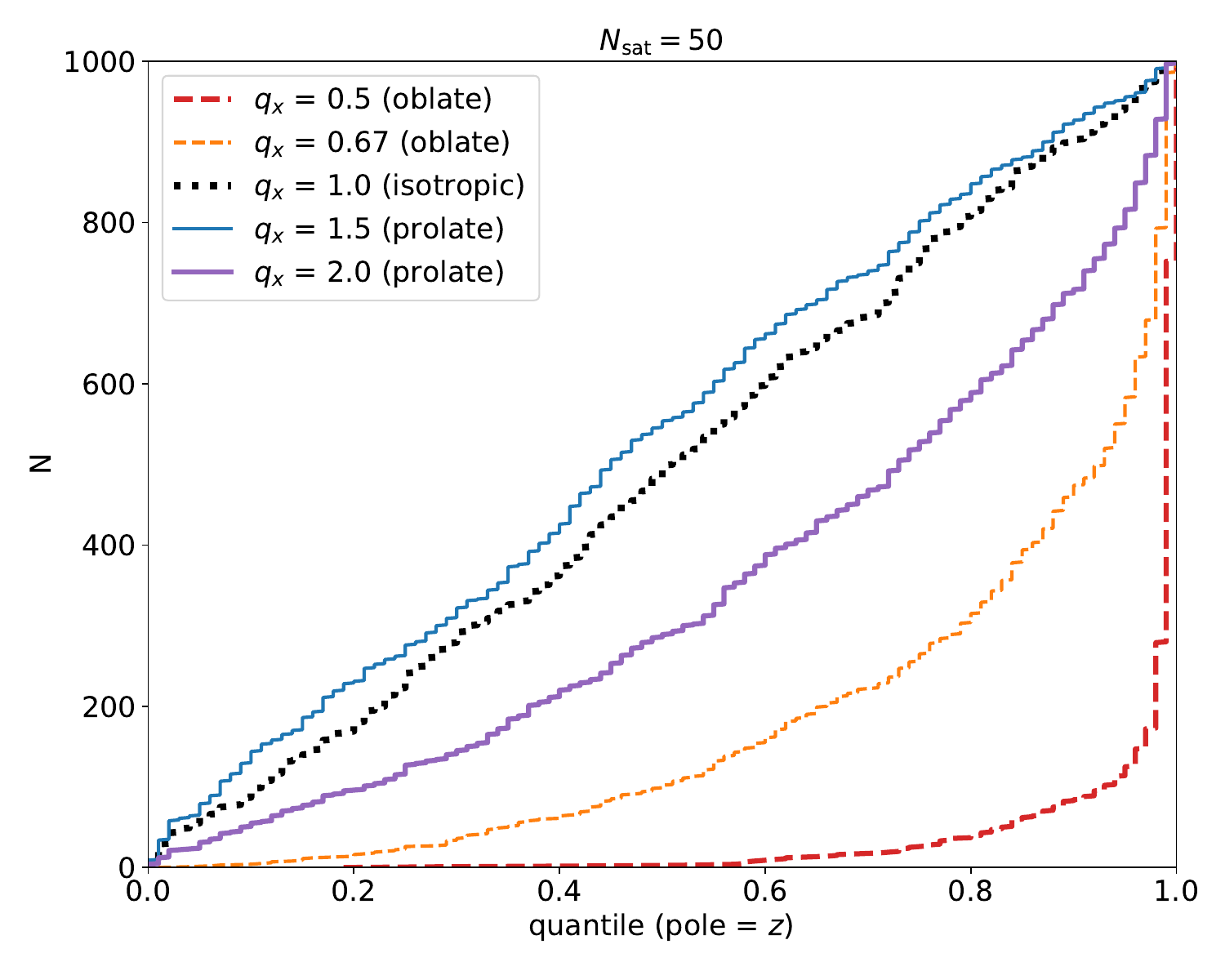}}
   \resizebox{\hsize}{!}
   {\includegraphics[width=0.32\hsize]{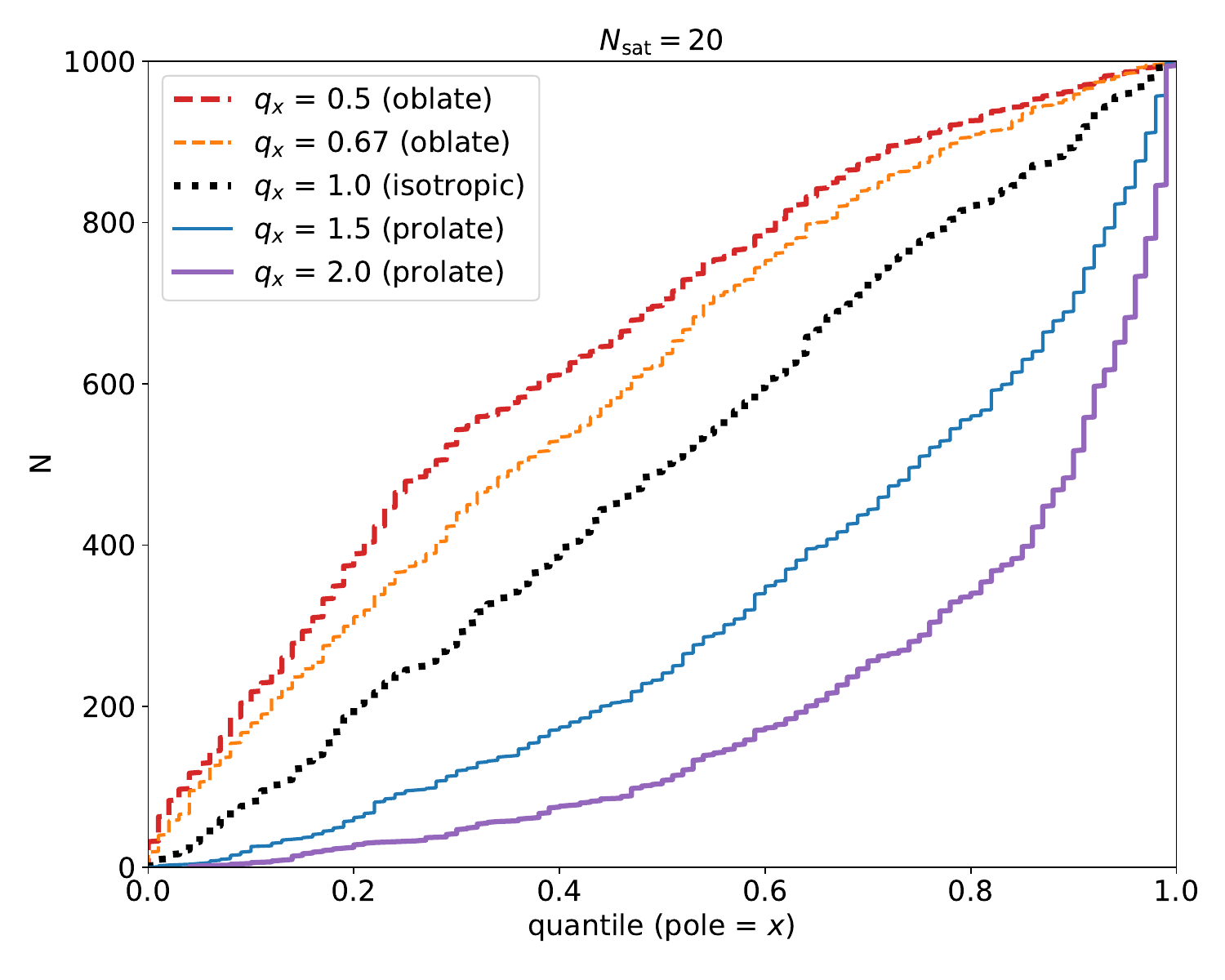}
   \includegraphics[width=0.32\hsize]{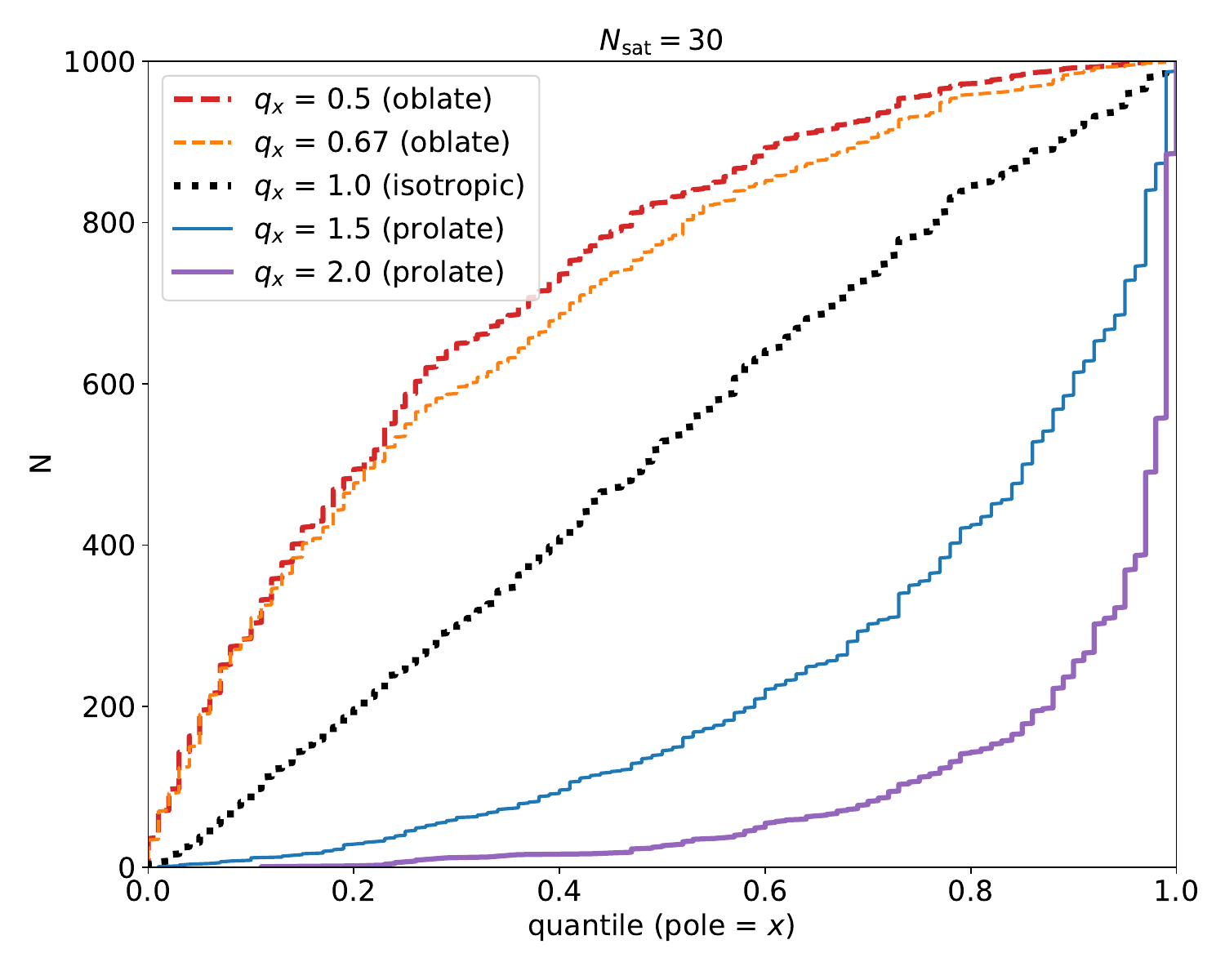}
   \includegraphics[width=0.32\hsize]{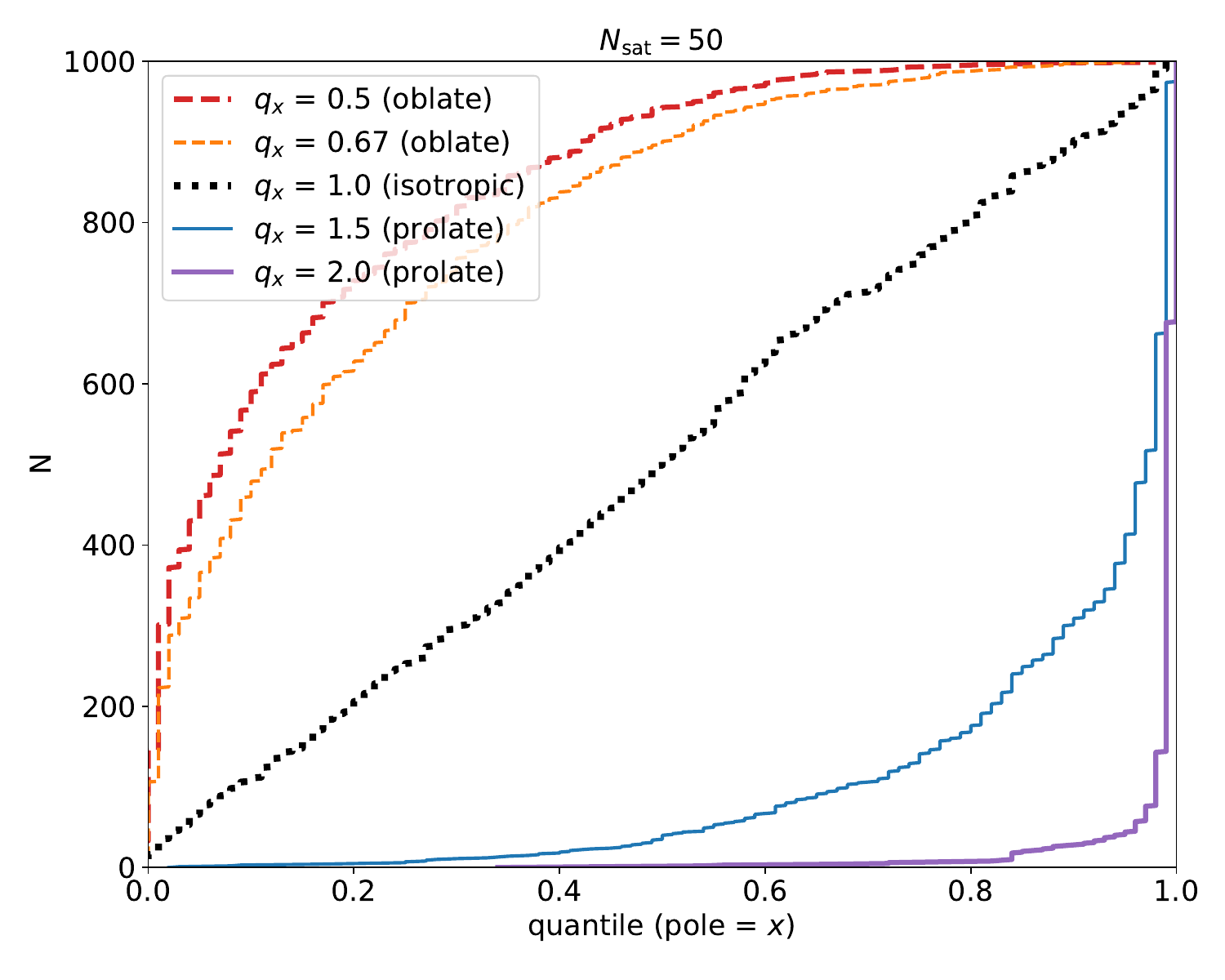}}
      \caption{Effect of the number of satellites on the inferred ''planarity'' of a distribution with intrinsic pro- or oblateness. From left to right the number of satellites per system is 20, 30, and 50, respectively (see Fig. \ref{fig:shape} for $N_\mathrm{sat}=40$). The upper panels orient the metric's pole along the $z$-, the lower panels along the $x$-axis.
      }
         \label{fig:numbers}
   \end{figure*}

\citet{2024arXiv241117813U} do not require that the number of satellites in the Milky Way sample is matched by its simulated analogs, but only that those analogs have $>30$\ satellites. While the metric determines the quantile of a given system relative to a sample of isotropic mock systems of the same number, the ''planarity'' of a system remains sensitive to the number of satellites considered. Any metric that refers to the likelihood that a given configuration appears among its isotropic counterparts needs to account for the fact that, given some degree of underlying anisotropy that the system of interest is drawn from, a larger population of satellites will result in a reduced impact of sampling variance and thus a lower likelihood to occur in isotropy.

We tested this by varying the number of satellites drawn from otherwise identical flattened and stretched distributions (as in Sect. \ref{sect:shape}). The results are shown in Fig. \ref{fig:numbers} for $N_\mathrm{sat} = [20, 30, 50]$. We identify a strong dependence on the number of satellites, with the same degree of underlying ob- or prolateness resulting in stronger effects on the quantile distribution for systems of larger $N_\mathrm{sat}$. This hinders comparability across different sample sizes, such as between the observed Milky Way satellites and simulated systems.

\subsection{Effect of satellite clustering}

   \begin{figure}[h!]
   \centering
   \includegraphics[width=0.9\hsize]{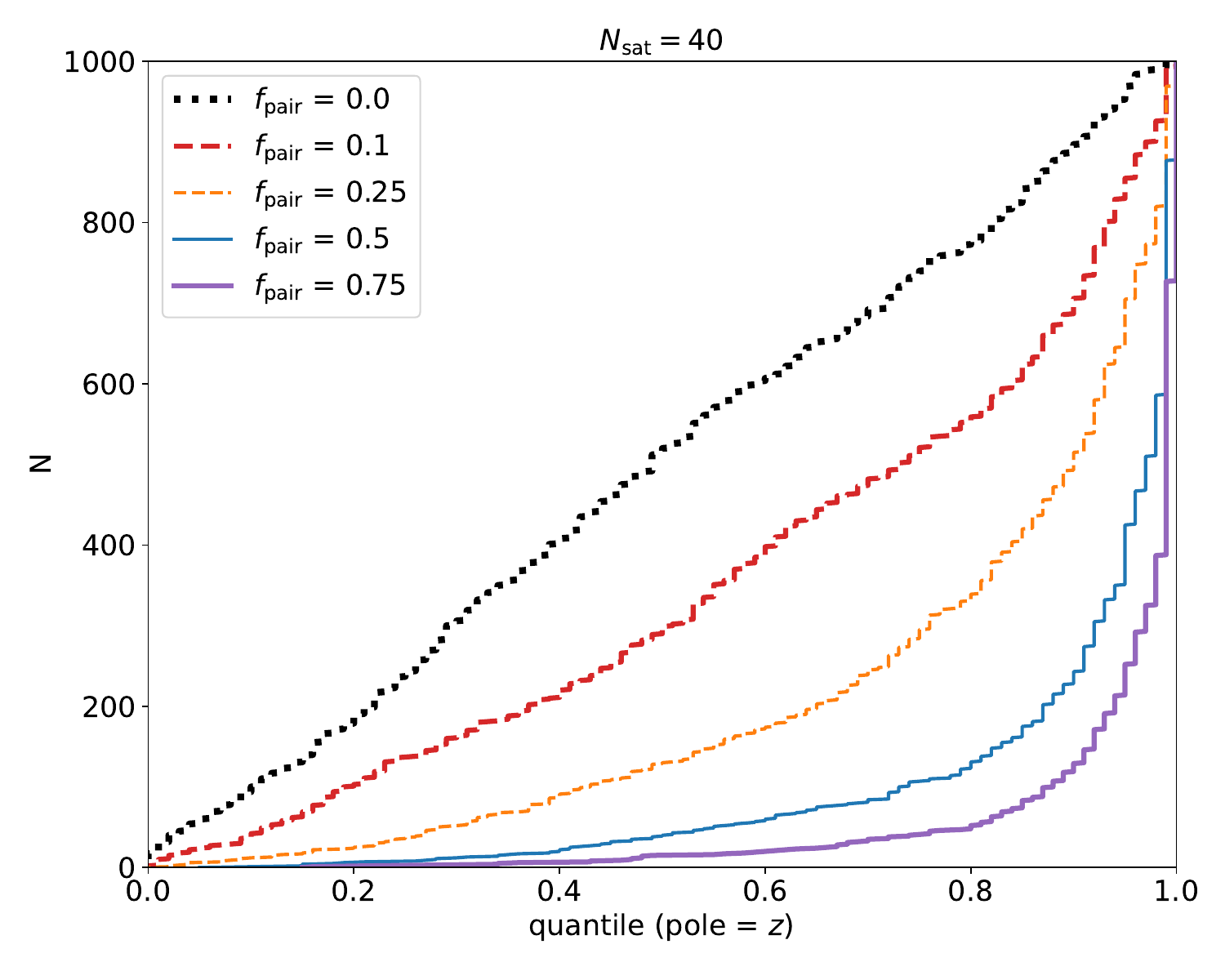}
      \caption{Distribution of quantiles for isotropic distributions with different fractions of paired satellites. Even a mild degree of clustering results in a substantial increase of high-quantile results.}
         \label{fig:pairs}
   \end{figure}

Galaxies in $\Lambda$CDM cluster hierarchically \citep{1978MNRAS.183..341W,1991ApJ...379...52W}. As such, we expect that at least some satellite galaxies have another dwarf companion nearby, be it a current or former satellite (\citealt{2015MNRAS.453.1305W, 2020MNRAS.495.2554E, 2020ApJ...893..121P, 2022ApJ...932...70P, 2023A&A...673A.160M, 2024MNRAS.527..437V}), a pair of dwarfs (\citealt{2014MNRAS.440.1225E, 2013MNRAS.431L..73F, 2014ApJ...795L..35C, 2018MNRAS.480.3376B, 2024ApJ...962..162C, 2024A&A...688A.153P}) or as part of an infalling group (\citealt{2013MNRAS.429.1502W, 2015ApJ...807...49W, 2024A&A...687A.212J}).

   \begin{figure*}[h!]
   \resizebox{\hsize}{!}
   {\includegraphics[width=0.32\hsize]{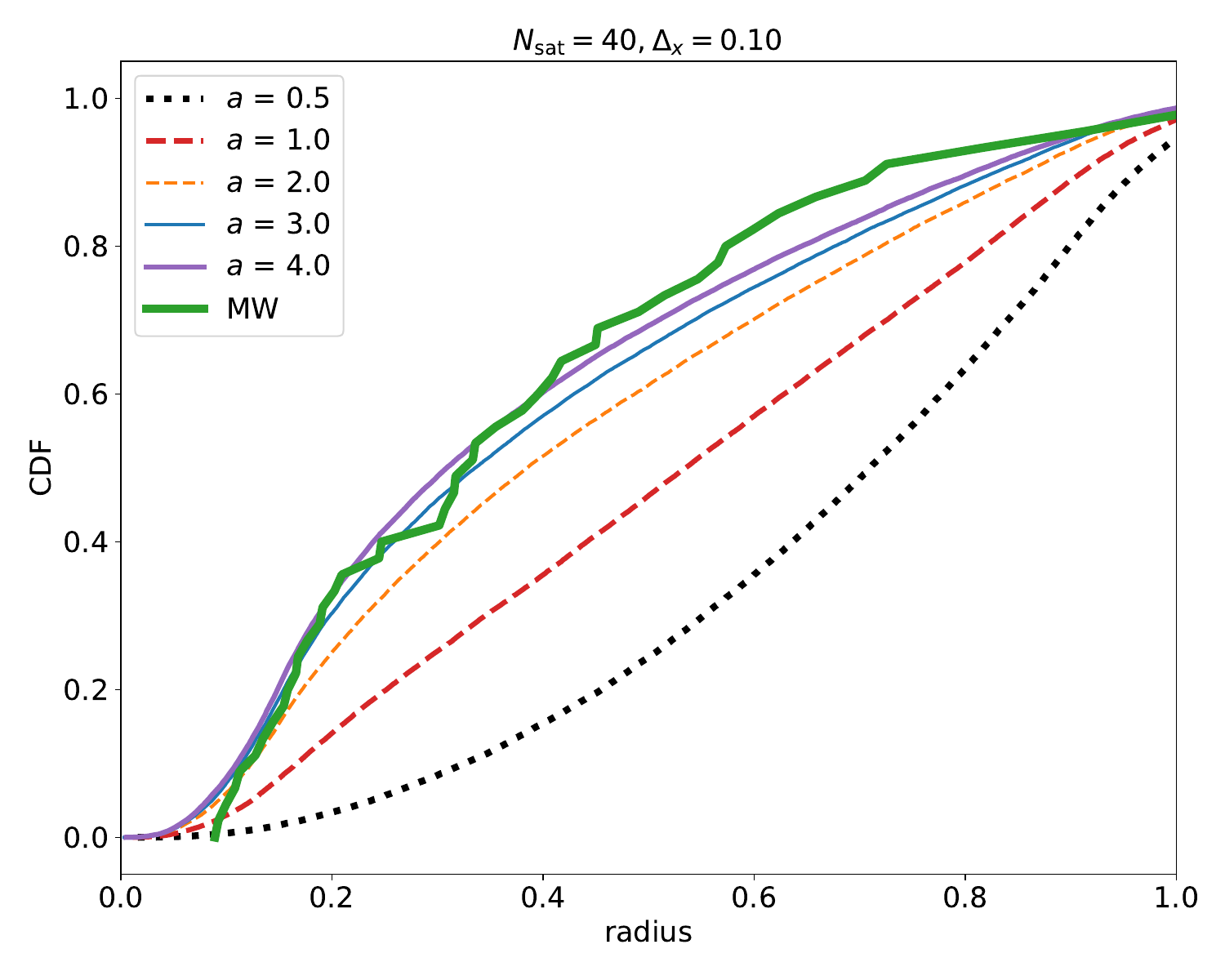}
   \includegraphics[width=0.32\hsize]{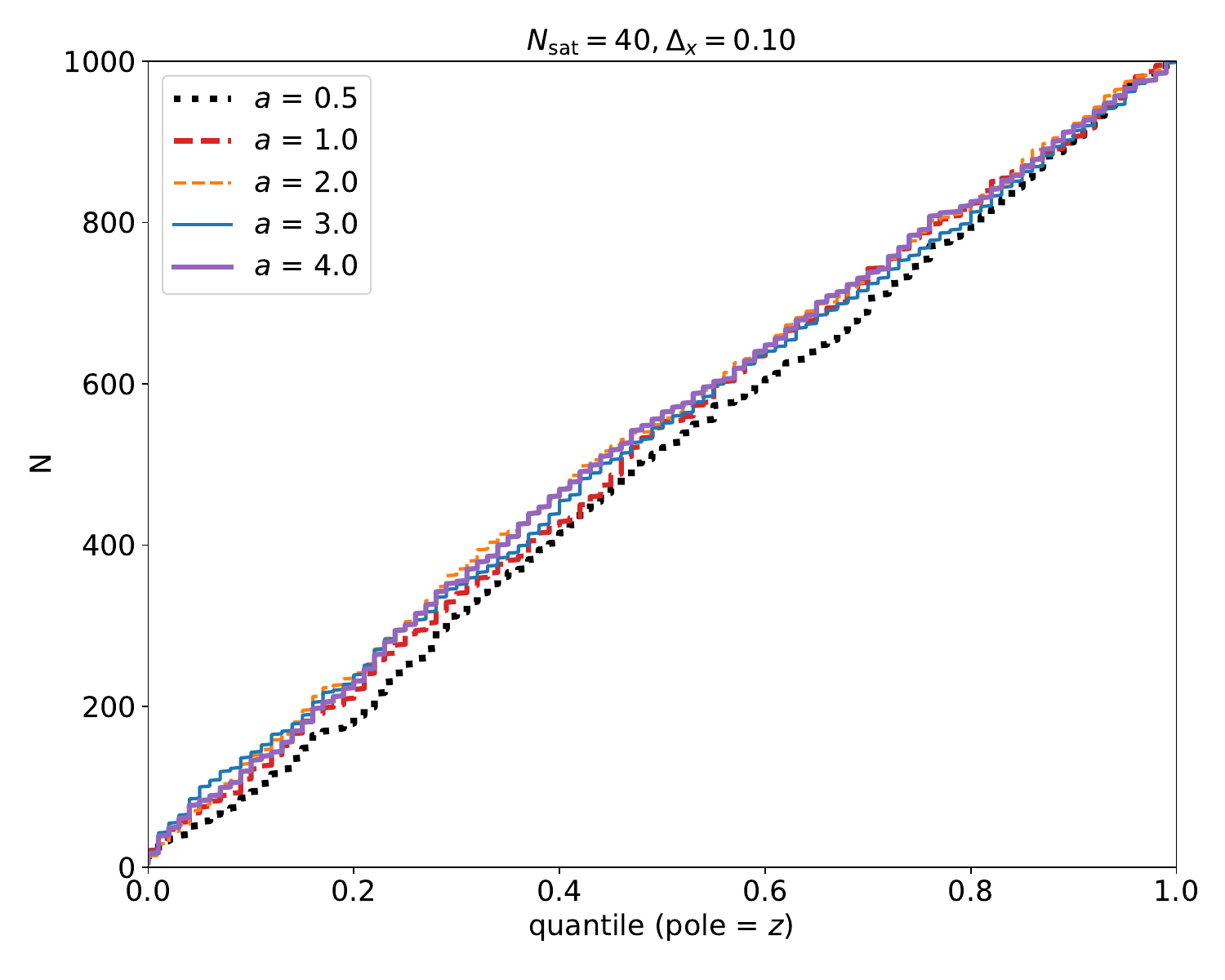}
   \includegraphics[width=0.32\hsize]{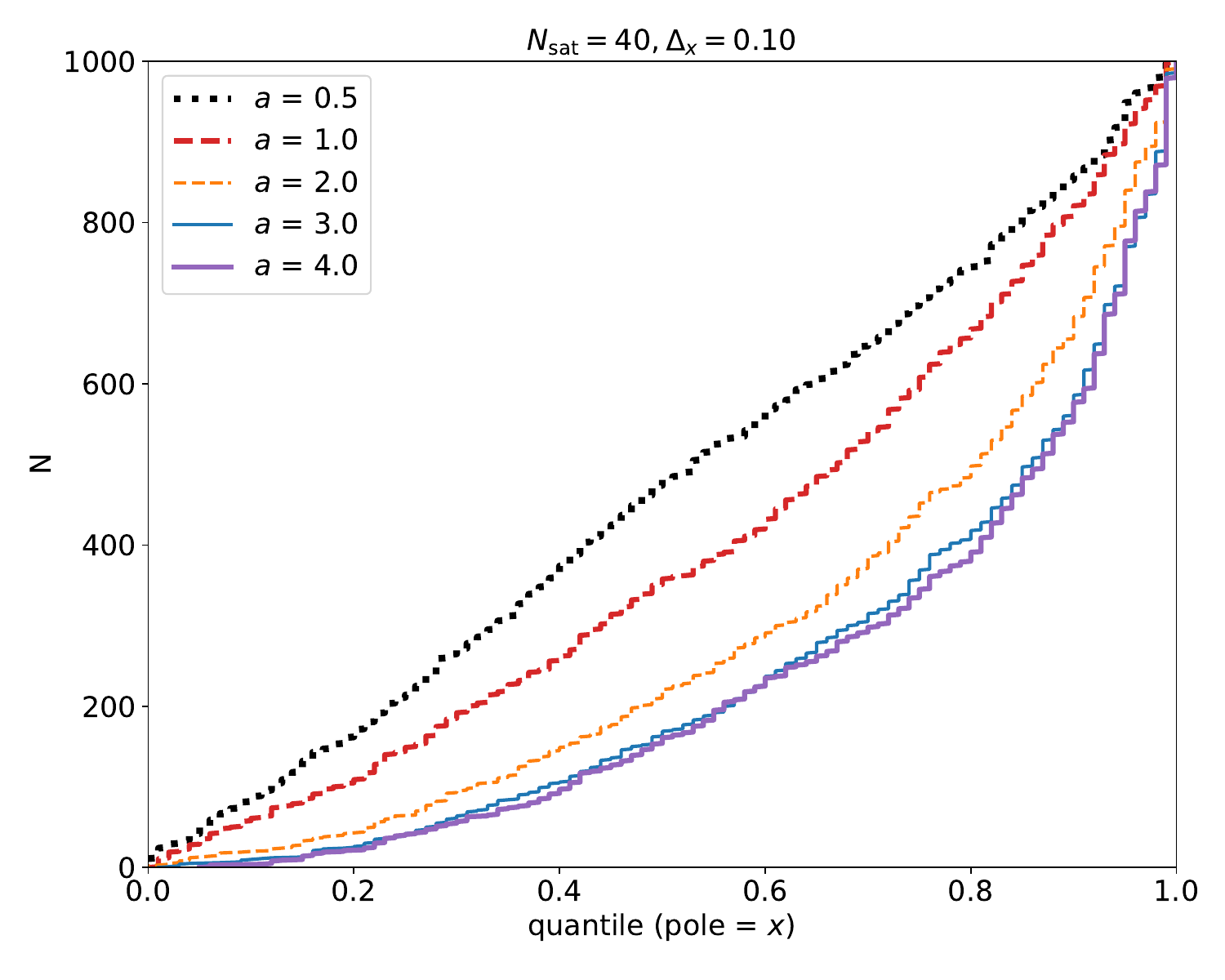}}
      \caption{Effect of radial distribution and lopsidedness on the inferred ''planarity'' of a satellite system.  
      The left panel plots the cumulative radial distribution of mock satellite systems (green: observed MW; black: fiducial distribution of \citealt{2024arXiv241117813U}).
      The middle (right) panel shows the quantile distribution when the metric's pole aligns with the $z$($x$)-axis and thus perpendicular to (along) the offset.
      }
         \label{fig:lopsidedness}
   \end{figure*}

Such clustering is independent of the presence of satellite planes. It is thus important to test whether satellite clustering affects the inferred planarity. We build a toy model to test this, by generating an isotropic system as in the fiducial case, but now each generated satellite has a chance $f_\mathrm{pair}$ to have a second satellite nearby. For each generated satellite we draw a random number from a uniform distribution between 0 and 1. If this is smaller than the probability $f_\mathrm{pair}$, then it is a primary of a pair and we add a secondary satellite nearby. We chose the secondary's position as an offset from the primary's position from a flat distribution in all three Cartesian directions, restricted to a maximum range of 10\% of the total extent of the system (0.1 for the adopted unit sphere). We stop the process once -- counting primaries and secondaries alike -- the total number of requested satellites is reached.

We generate 1000 such systems per $f_\mathrm{pair} = [0.0, 0.1, 0.25, 0.5, 0.75]$, apply the ''planarity'' metric, and record the resulting quantiles. Fig. \ref{fig:pairs} visualizes the results. Even a mild degree of clustering (one in ten primary satellites gets assigned a secondary) leads to a substantial increase in the number of high-quantile cases.
Clustering in a galaxy system, a natural occurrence in $\Lambda$CDM's hierarchical formation scenario, thus also introduces a strong bias to infer higher degrees of ''planarity'' with the proposed metric, even in the absence of an underlying satellite plane.

\subsection{Effect of asymmetry/lopsidedness}

Both observed and simulated systems show radial distributions with larger satellite densities in the inner than the outer regions \citep{2010MNRAS.402.1995M, 2019MNRAS.487.4409K, 2020MNRAS.491.1471S}. In addition, satellite systems show asymmetries, most prominently an overall lopsidedness with more satellites on one side of their host than the other \citep{2013ApJ...766..120C, 2016ApJ...830..121L, 2020ApJ...898L..15B, 2022ApJ...938..101S, 2024A&A...690A.110H}. Similar features are present in satellite systems in cosmological simulations \citep{2017ApJ...850..132P, 2021ApJ...914...78W, 2023ApJ...947...56S, 2024MNRAS.529.1405L}. Such lopsidedness does not constitute a plane-like satellite distribution, and thus a metric of planarity should not be sensitive to overall asymmetries or shifts in satellite distributions.

To test this we set up model satellite distributions offset from the host and with different radial distributions, as shown in the left panel of Fig. \ref{fig:lopsidedness}. The fiducial radial distribution of \citet{2024arXiv241117813U} (black dotted line) assumes a uniform spatial density and results in considerably more spread-out satellite distributions than the observed Milky Way system (green line, data from \citep{2021ApJ...916....8L} normalized to the most distant considered satellite Leo\,I). 

We generate different distributions by drawing the radial distance of each model satellite from a flat distribution in $r' = [0, 1]$, and then assign it a radius $r = r'^a$, with an exponent of $a$\ between $a = 0.5$ (the fiducial radial distribution of \citealt{2024arXiv241117813U}), to $a = 4$ (a highly concentrated distribution).
These systems are set up isotropically, and shifted by 10\% of their maximum extent along the $x$-axis (0.1 for the adopted unit sphere). To make them align better with the observed Milky Way satellite system, we reject satellites within the inner 5\% of radius, which would be close to or within the Galactic disk.

The middle and right panels of Fig. \ref{fig:lopsidedness} show the resulting quantile distributions. When the metric's pole points along the $z$-axis (perpendicular to the offset), no impact on the quantiles is apparent. However, when the pole aligns with the direction of the offset (the $x$-axis) then the quantiles are biased to higher values. This effect is stronger for more concentrated distributions. Thus, a small asymmetric offset in the satellites and a realistic radially concentrated satellite distribution -- natural occurrences in $\Lambda$CDM independent of satellite planes -- also bias to high inferred quantiles and can thus let one to falsely infer a higher degree of planarity when employing this metric.


\section{Conclusions}

We have investigated the behavior and properties of the new ''planarity'' metric proposed by \citet{2024arXiv241117813U} and used to claim consistency between the Milky Way's plane of satellite galaxies and $\Lambda$CDM simulations. We find that the metric's results are sensitive to the chosen orientation of its spherical coordinate system, with resulting quantile values of mock systems rotated by $90^\circ$\ showing almost no correlation. This property alone makes it an inadequate tool to measure, infer or compare satellite galaxy systems. 

Furthermore, we have tested the metric's response to other types of phase-space correlation present in satellite galaxy systems. We find that the overall deviation of the shape from sphericity, satellite clustering, and lopsided satellite distributions, can all result in the proposed ''planarity'' metric returning high quantile values for mock systems. These features of the metric developed by \citet{2024arXiv241117813U} will overestimate the inferred occurrence of planes in cosmological simulations. Since all of these effects are present in $\Lambda$CDM satellite systems, but independent of the presence of satellite planes, the metric cannot be used to infer consistency of the Milky Way satellite {\it plane}  with $\Lambda$CDM.

Taken together, we find the proposed ''planarity'' metric to be unreliable due to its sensitivity to the satellite system's orientation, biased to return inflated degrees of apparent ''planarity'' in the presence of other types of phase-space correlations resulting in a lack of specificity, and thus overall inadequate to study the planes of satellite galaxies issue.


\begin{acknowledgements}
Marcel S. Pawlowski acknowledges funding via a Leibniz-Junior Research Group (project number J94/2020). O.M. is grateful to the Swiss National Science Foundation for financial support under the grant number PZ00P2\_202104.
\end{acknowledgements}

\bibliographystyle{aa}
\bibliography{planarity.bib}

\appendix

\section{Error Treatment}
\label{appendix}

In analyzing the observed Milky Way system of satellite galaxies, \citet{2024arXiv241117813U} employ a Monte-Carlo sampling scheme to account for measurement errors. From the measured positions, distances, line-of-sight velocities, and proper motions and their errors, the resulting 6D Cartesian coordinates of the satellite galaxies and their spread due to measurement errors have been obtained. \citet{2024arXiv241117813U} then sample from these distributions, drawing from each Cartesian coordinate independently. 

This neglects the presence of strong correlations in the possible positions and velocities of a satellite galaxy, which generally do not align with the axes of the Galactic Cartesian coordinate system. Sampling the Cartesian coordinates independently thus results in incorrect realizations which are physically impossible given the measured constraints on the observed satellite galaxies.

We demonstrated this by following the same procedure. We generate Monte-Carlo realizations of a given satellite galaxy by drawing from its position, distance, line-of-sight velocity, and proper motion errors (for simplicity assumed to be normal distributed), using data from \citet{2022A&A...657A..54B}. These realizations are then converted to Galactic Cartesian coordinates. We measure the median position and standard deviation in each of these coordinates. These are then used as input for a second round of Monte-Carlo realizations, where we now follow the procedure used by \citet{2024arXiv241117813U} and treat each Cartesian coordinate independently. The result is a sample of realizations in Cartesian space, which we convert back to spherical Galactic coordinates. In other words, for each realization we calculate the resulting position, distance, line-of-sight velocity, and proper motion.

   \begin{figure}[h!]
   \centering
   \includegraphics[width=\hsize]{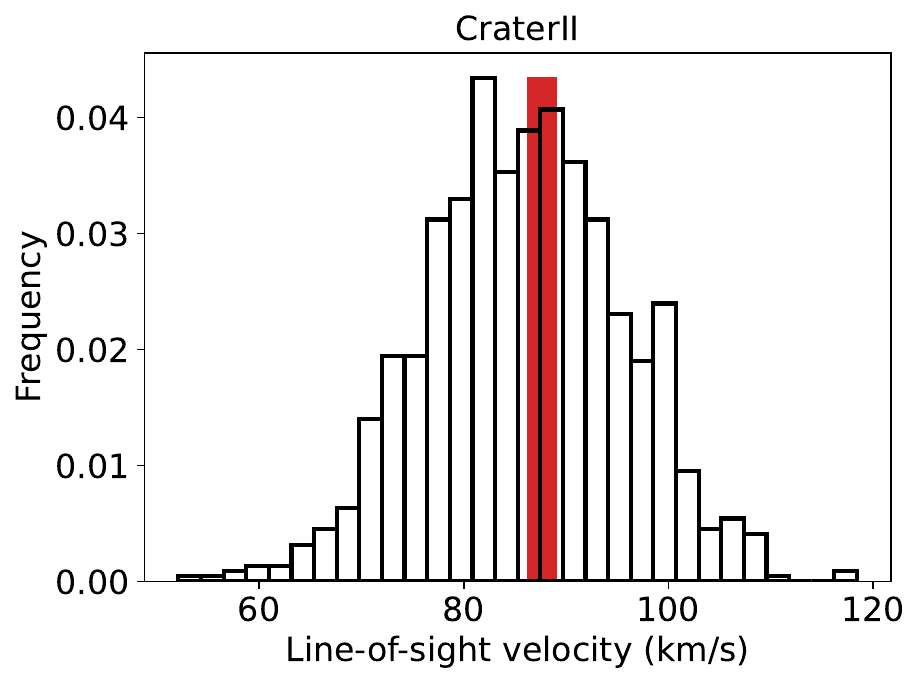}
   \includegraphics[width=\hsize]{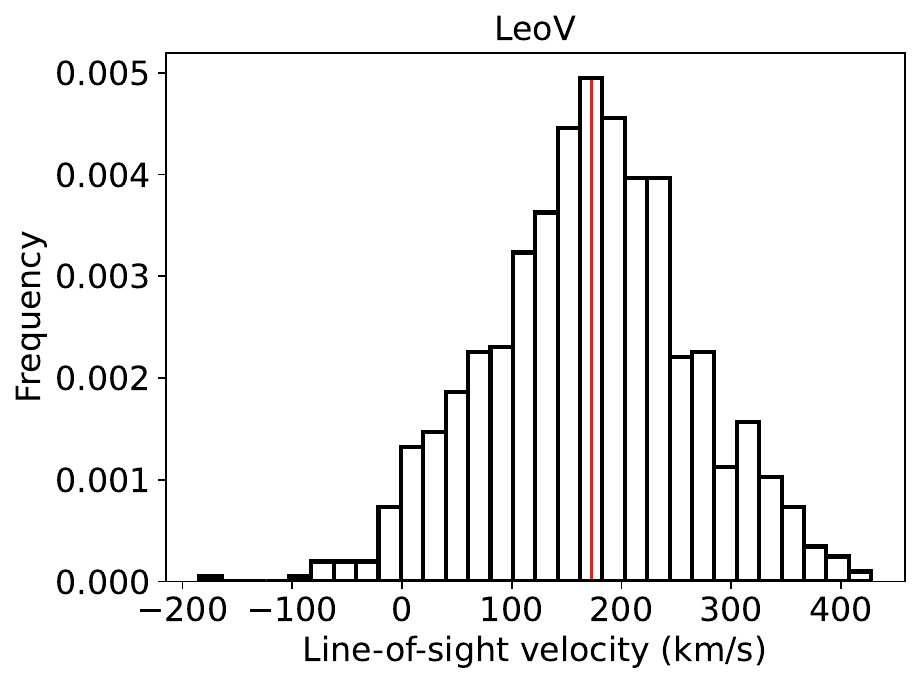}
      \caption{Distributions of line-of-sight velocities from Monte-Carlo sampling for the two Milky Way satellite galaxies Crater\,II (upper panel) and Leo\,V (lower panel). The red bands give the range from sampling the measured line-of-sight velocities. The black histograms show the resulting line-of-sight velocities if errors are sampled in 6D Cartesian coordinates independently, ignoring their mutual correlations. 
      }
         \label{fig:error}
   \end{figure}

If the procedure were correct, the resulting Galactic coordinates and their spread should match with the measurement constraints. In Fig.  \ref{fig:error}, we use observationally well constrained line-of-sight velocities (errors typically do not exceed a few km/s) to demonstrate that this is not the case.
The plots demonstrate the impact of incorrectly sampling measurement errors in the phase-space coordinates of Milky Way satellite galaxies. The red bands give the measured line-of-sight velocities to a satellite galaxy, their widths indicate the maximum extent from our original round of Monte-Carlo samplings. The black histogram shows the resulting line-of-sight velocity after converting the second round of Monte-Carlo sampling back to Galactic coordinates. Clearly the realizations sampled from the Cartesian distributions without considering their inherent correlations result in nonphysical phase-space positions.
    
Specifically, the upper panel in Fig. \ref{fig:error} shows the results for Crater\,II, a satellite galaxy with relatively well constrained proper motions. Even in this case the incorrect sampling results in line-of-sight velocities deviating substantially from the actually measured value by tens of km/s. The situation is much worse for satellites with only poorly constrained proper motions (of which there are many), as demonstrated in the lower panel using Leo\,V as an example. In this case, the incorrect error sampling results in an extremely wide spread of line-of-sight velocities, some offset by hundreds of km/s from the actual measured value.

\section{Scaled metric with equal-area bins}
\label{appendix2}

   \begin{figure}[h!]
   \centering
   \includegraphics[width=0.9\hsize]{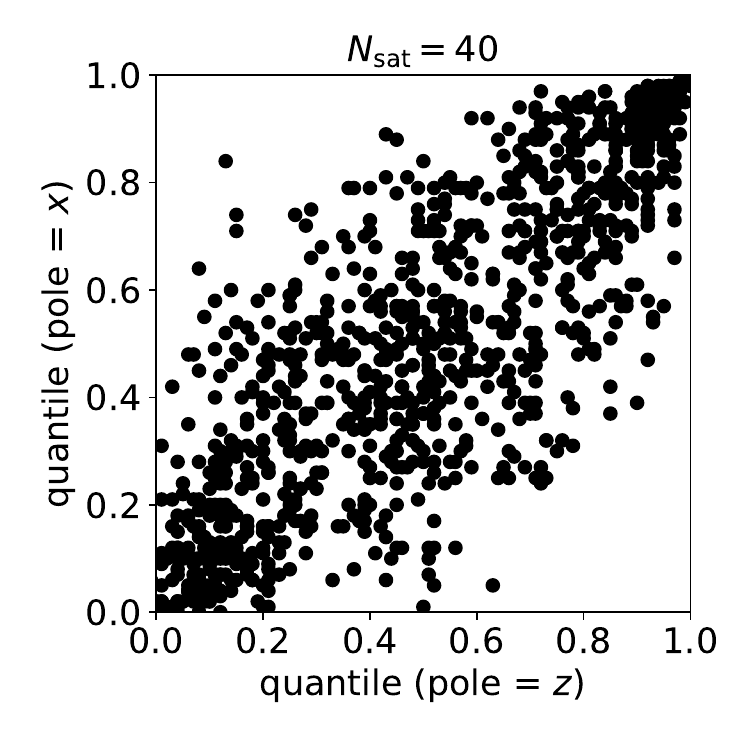}
      \caption{Quantiles for 1000 random isotropic distributions in two orientations rotated by $90^\circ$, like Fig. \ref{fig:orientation} but now with bins of equal area.}
         \label{fig:orientation_scaled}
   \end{figure}
   
We here summarize our results after ensuring that the bins of the underlying 2D histograms in the metric's spherical coordinates are of equal areas. To do this, we scale the original inclination angles $\beta$\ (that runs from 0 to $\pi$) to a new $\beta' = 0.5\ \pi\ (1 - \cos{\beta})$. This ensures that the histograms have the same axis ranges as shown in \citet{2024arXiv241117813U}. We emphasize that even with equal areas, the shapes of the bins differ, with the bins closer to the poles being more elongated than those close to the equator of the coordinate system. This suggests that the scaled metric's results remain sensitive to the orientation under which a satellite system is studied.

Figure \ref{fig:orientation_scaled} shows the test on the effect of orientation, and confirms that an effect remains even for the scaled metric. The scaling improves the situation somewhat, with more of a correlation apparent between the rotated test systems. However, there is still substantial scatter in the inferred degree planarity for different orientations, making the metric unreliable.

   \begin{figure}[h!]
   \centering
   \includegraphics[width=0.9\hsize]{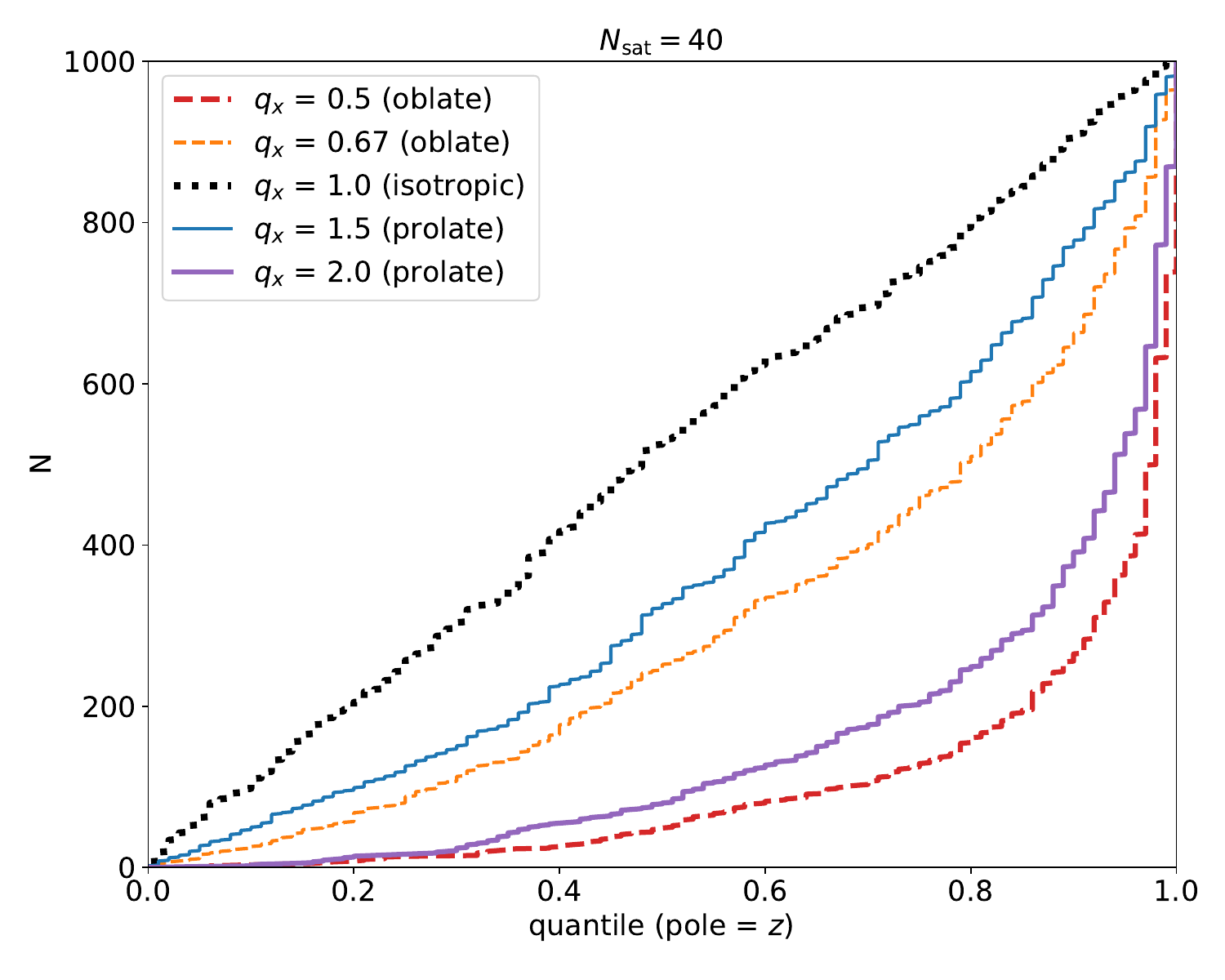}
   \includegraphics[width=0.9\hsize]{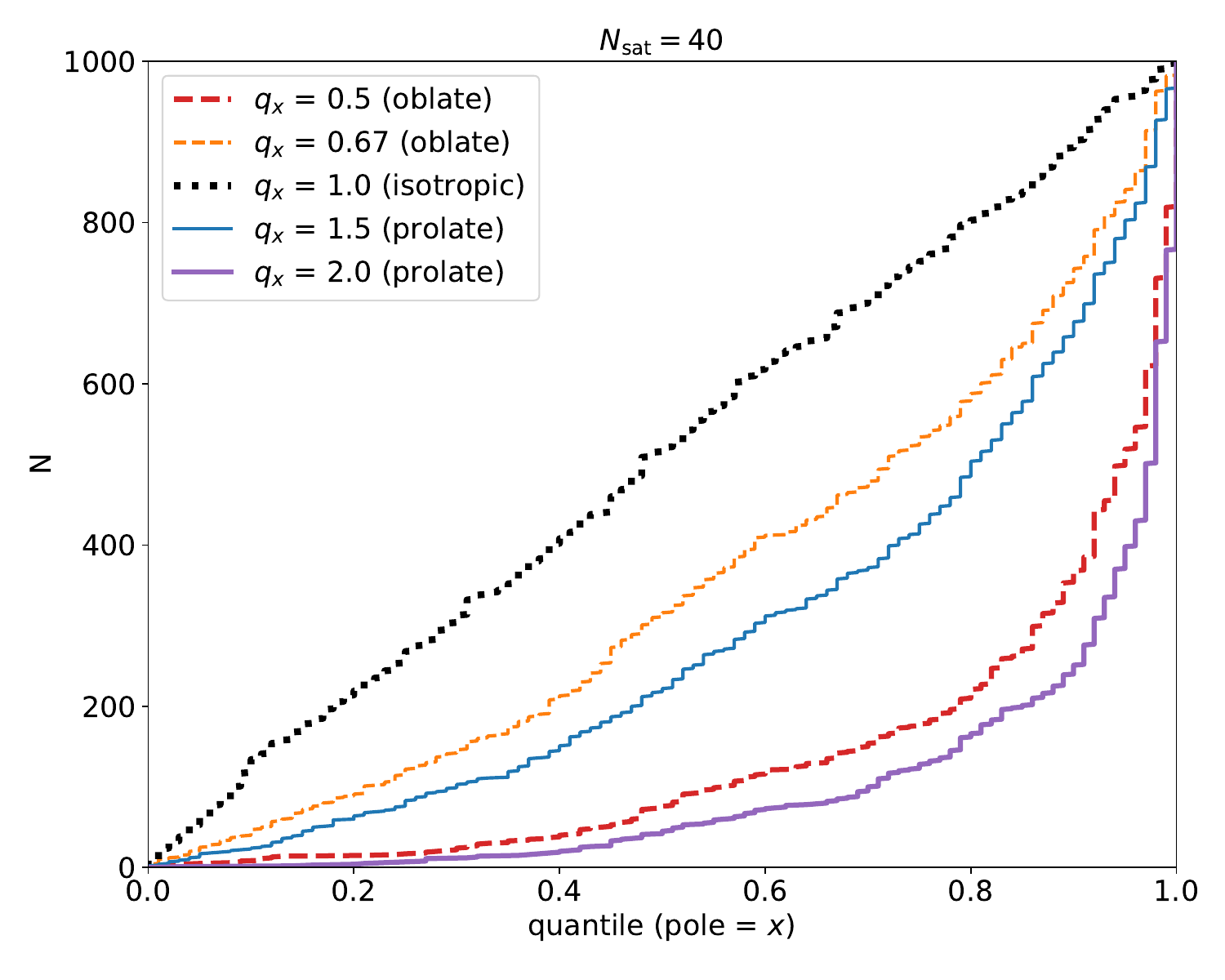}
      \caption{Distribution of quantiles in the scaled metric for systems with different degrees of flattening ($q < 1.0$, oblate) or elongation ($q > 1.0$, prolate) along the $x$-axis.
      For the upper panel the metric's pole is oriented along the $z$-, for the lower panel along the $x$-axis.
      }
         \label{fig:shape_scaled}
   \end{figure}
   
Figure \ref{fig:shape_scaled} repeats the test for prolate and oblate distributions. With the scaled metric the overall effect remains present, and in fact both prolate and oblate distributions now result in inferring increased degrees of planarity irrespective of the orientation. There does, however, remain some dependence on the orientation as can be seen be the lines for $q = 0.5$ and 2.0, as well as $q = 0.67$ and 1.5 effectively swapping placed between the upper and lower panels.

   \begin{figure*}[h!]
   \resizebox{\hsize}{!}
   {\includegraphics[width=0.32\hsize]{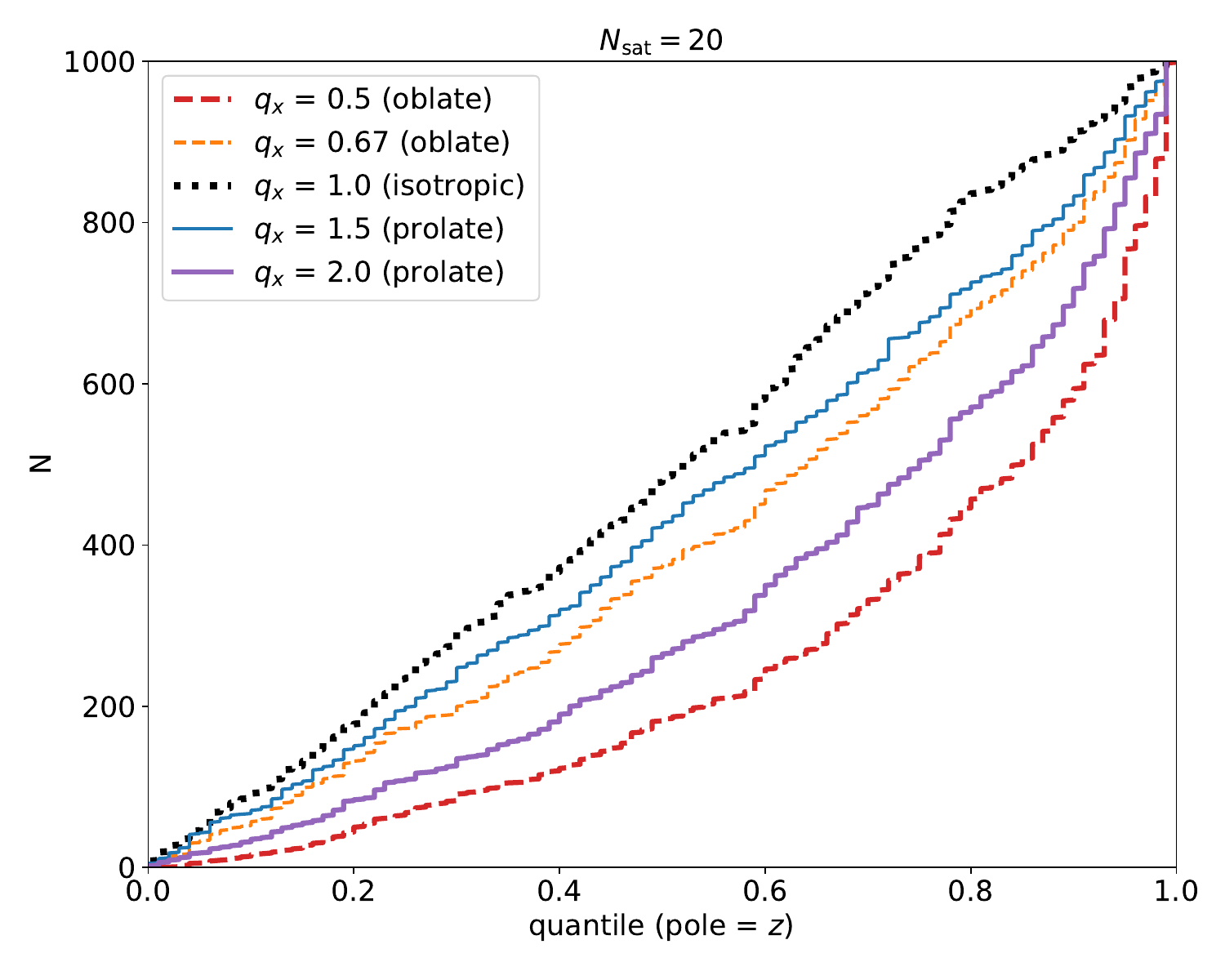}
   \includegraphics[width=0.32\hsize]{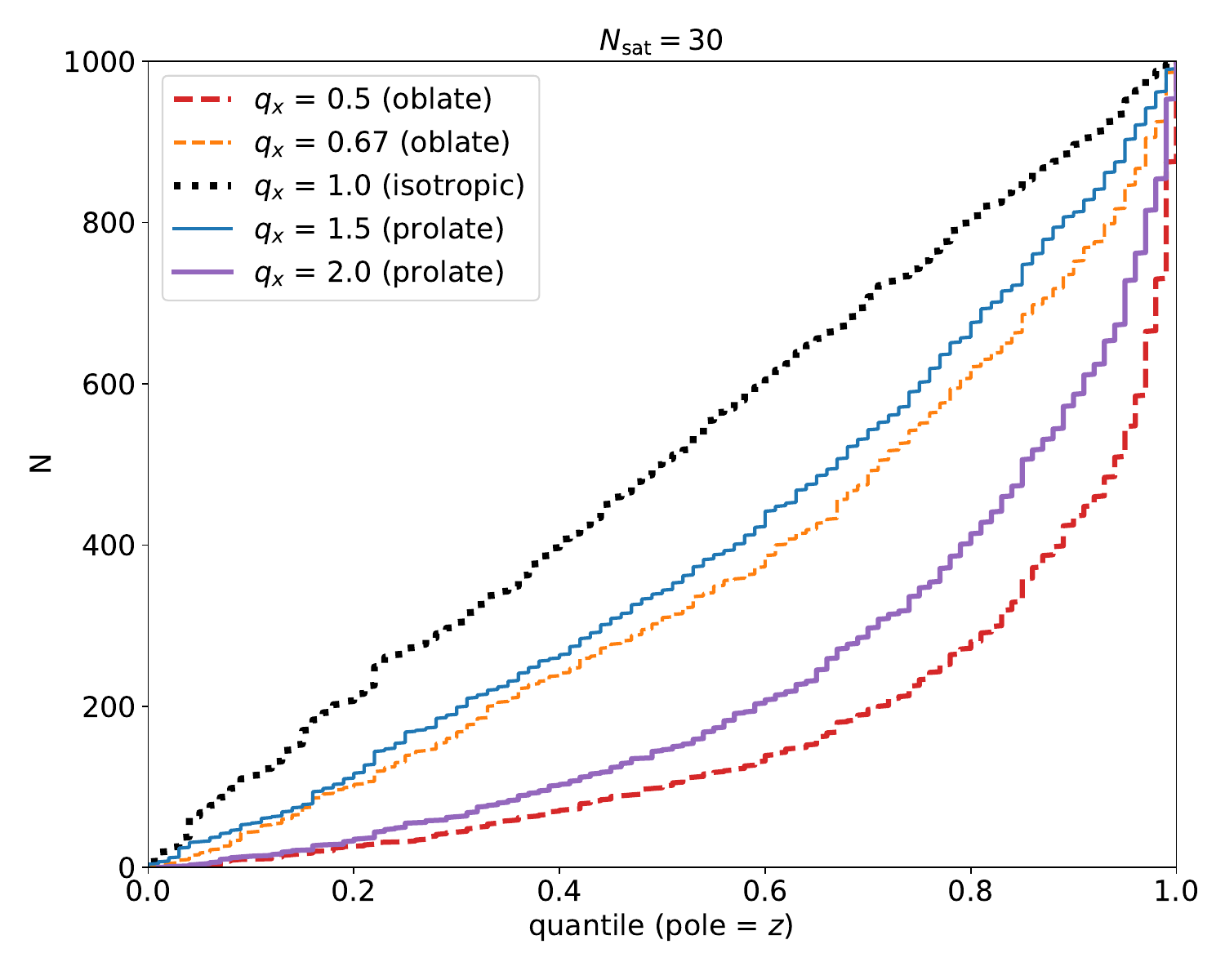}
   \includegraphics[width=0.32\hsize]{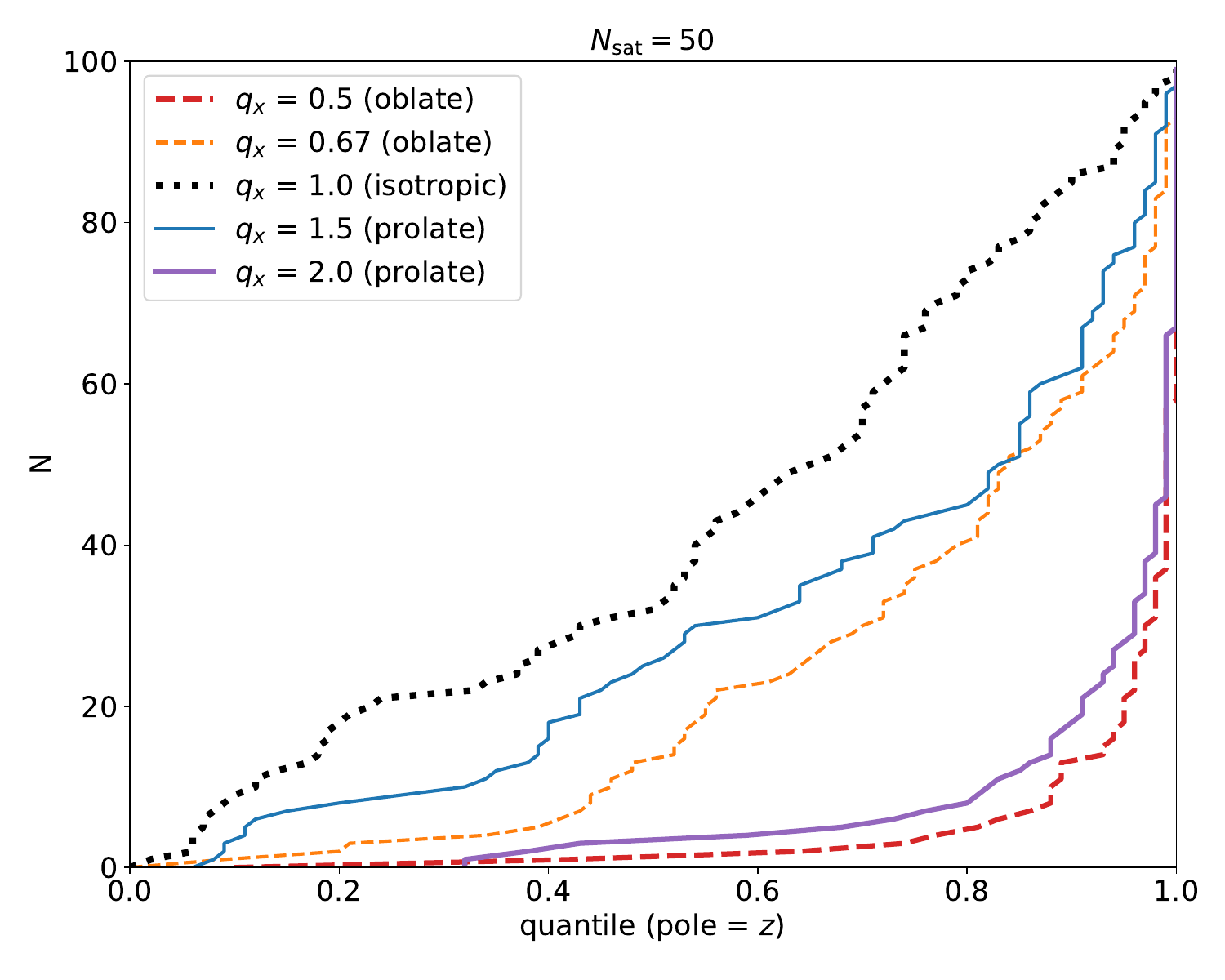}}
   \resizebox{\hsize}{!}
   {\includegraphics[width=0.32\hsize]{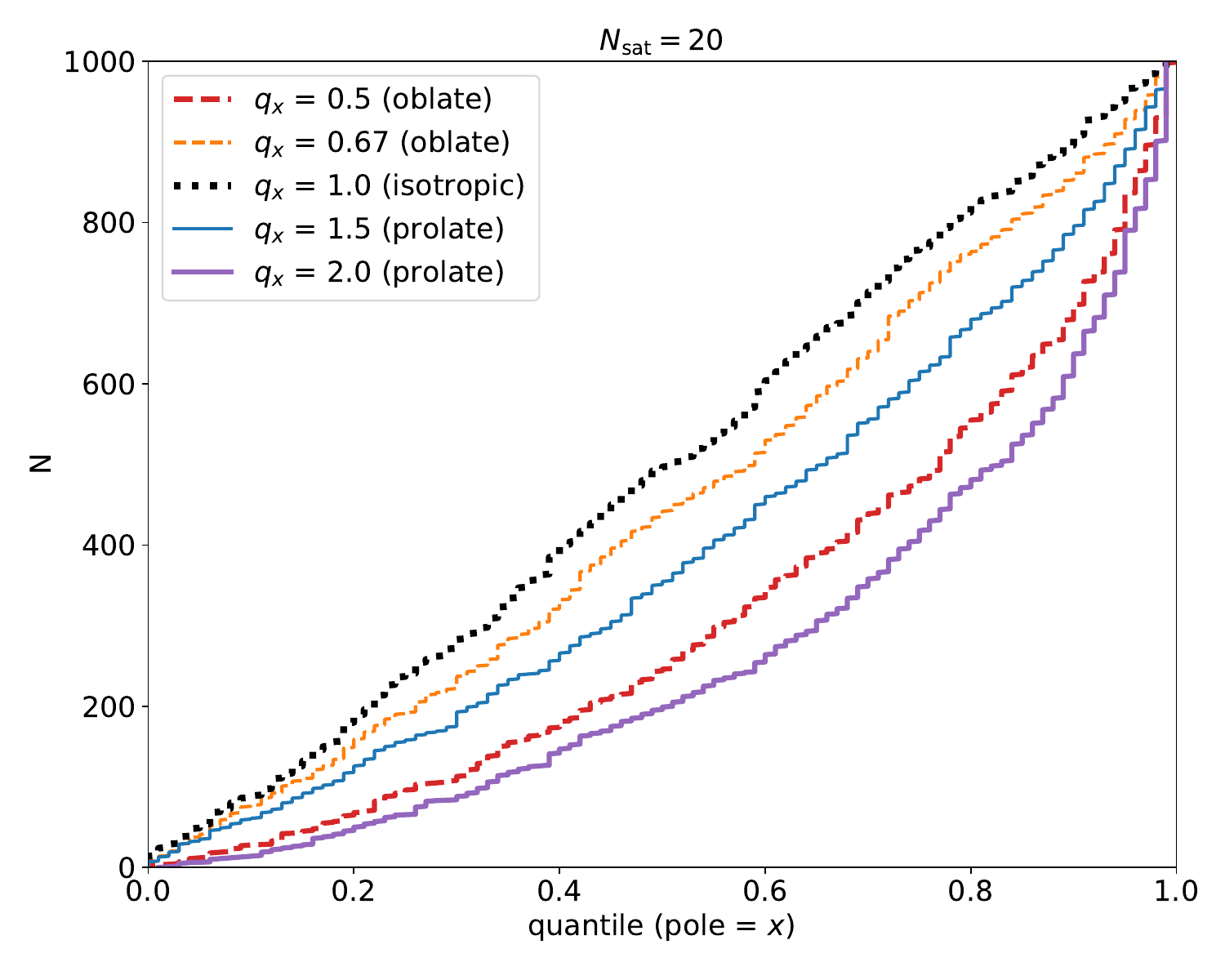}
   \includegraphics[width=0.32\hsize]{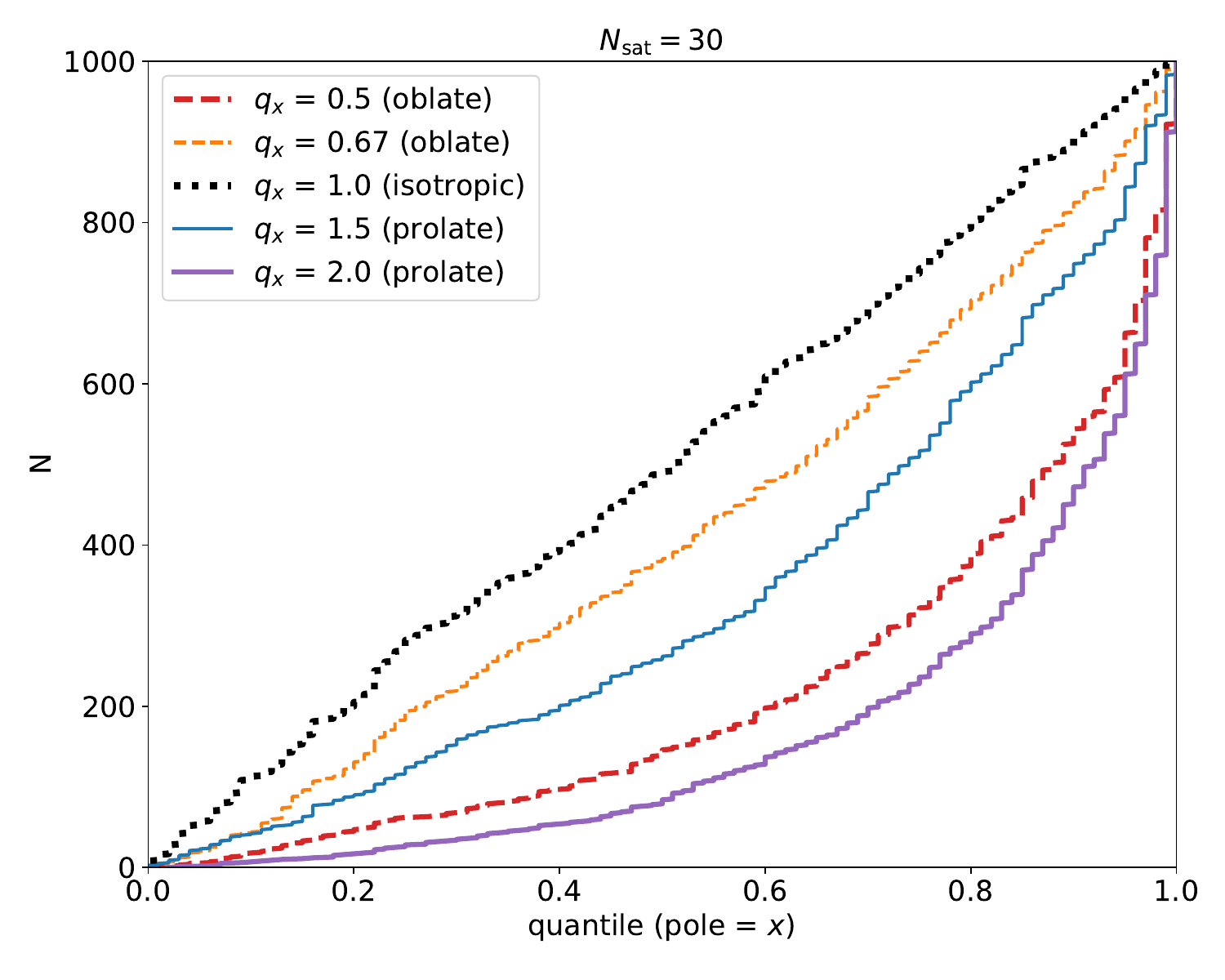}
   \includegraphics[width=0.32\hsize]{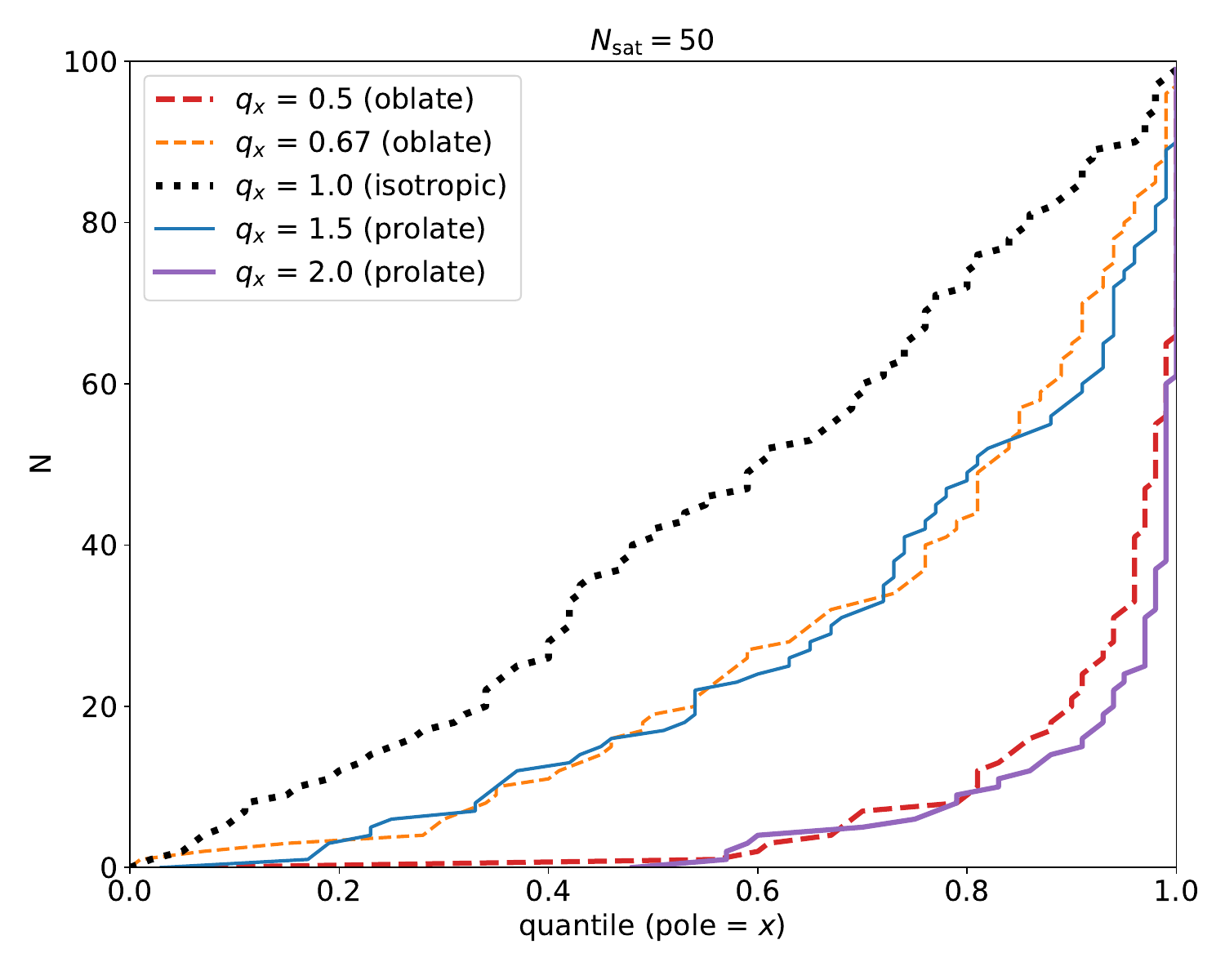}}
      \caption{Effect of the number of satellites on the inferred ''planarity'' of a distribution with intrinsic pro- or oblateness, using the scaled metric.
      }
         \label{fig:numbers_scaled}
   \end{figure*}

Figure \ref{fig:numbers_scaled} shows that also in case of the scaled metric, the number of satellites in a system has an influence on the inferred degree of ''planarity'' if the systems are drawn from pro- or oblate distributions.

   \begin{figure}[h!]
   \centering
   \includegraphics[width=0.9\hsize]{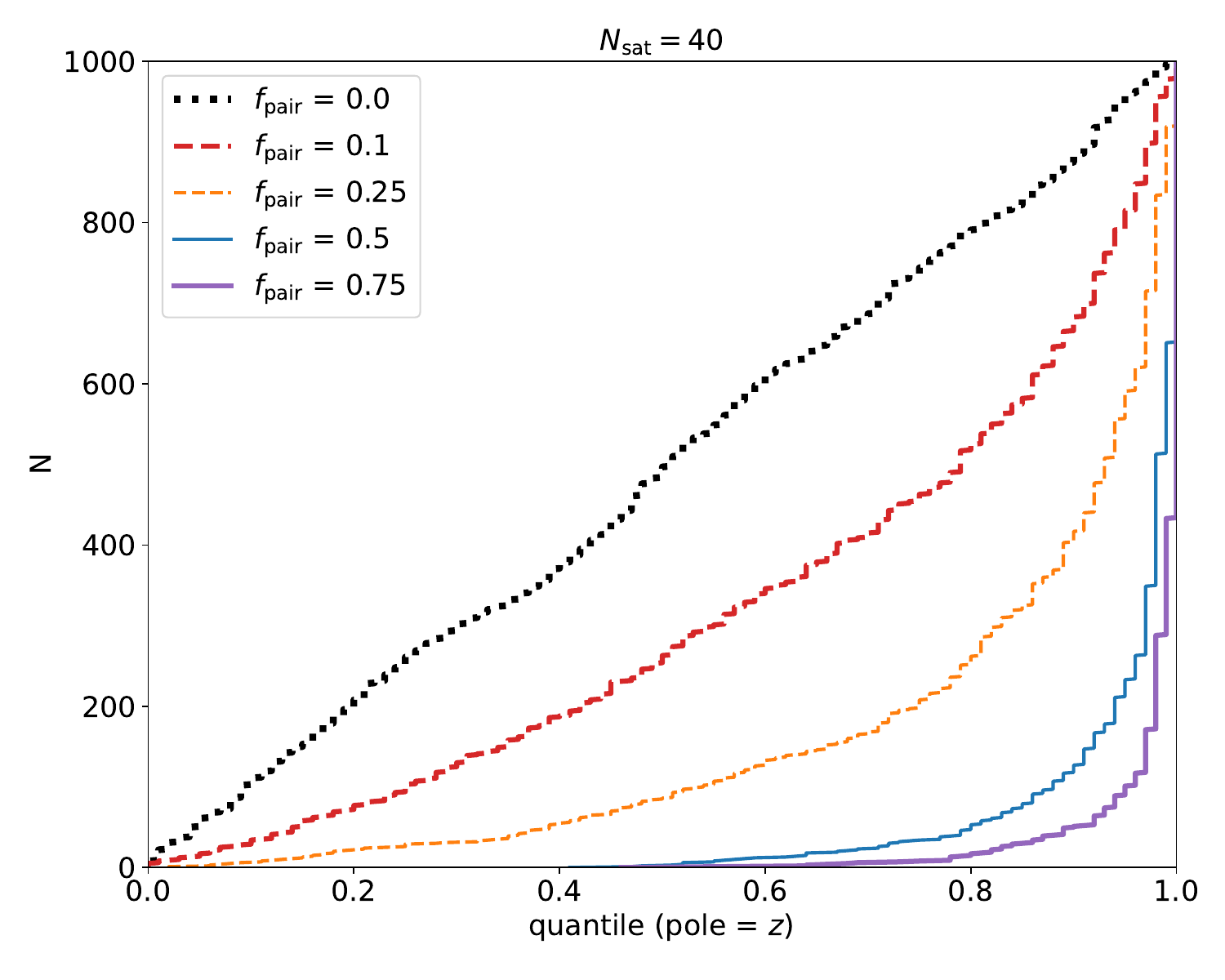}
      \caption{Distribution of quantiles in the scaled metric for isotropic distributions with different fractions of paired satellites.}
         \label{fig:pairs_scaled}
   \end{figure}

Figure \ref{fig:pairs_scaled} shows that the scaled metric also returns increased degrees of ''planarity'' if satellite galaxies show clustering modeled as satellite pairs, more so than in the non-scaled case. 
   
   \begin{figure*}[h!]
   \resizebox{\hsize}{!}
   {
   \includegraphics[width=0.45\hsize]{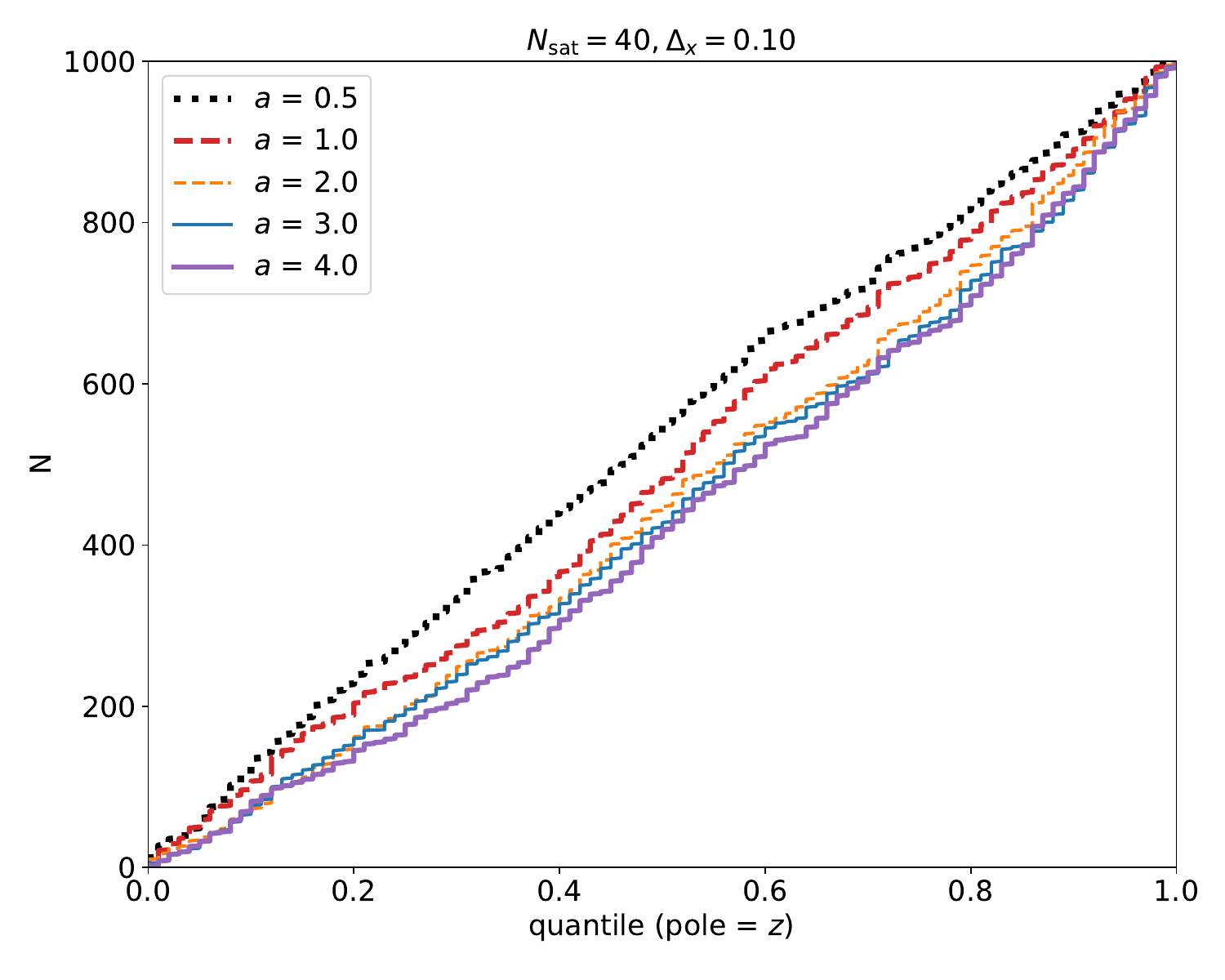}
   \includegraphics[width=0.45\hsize]{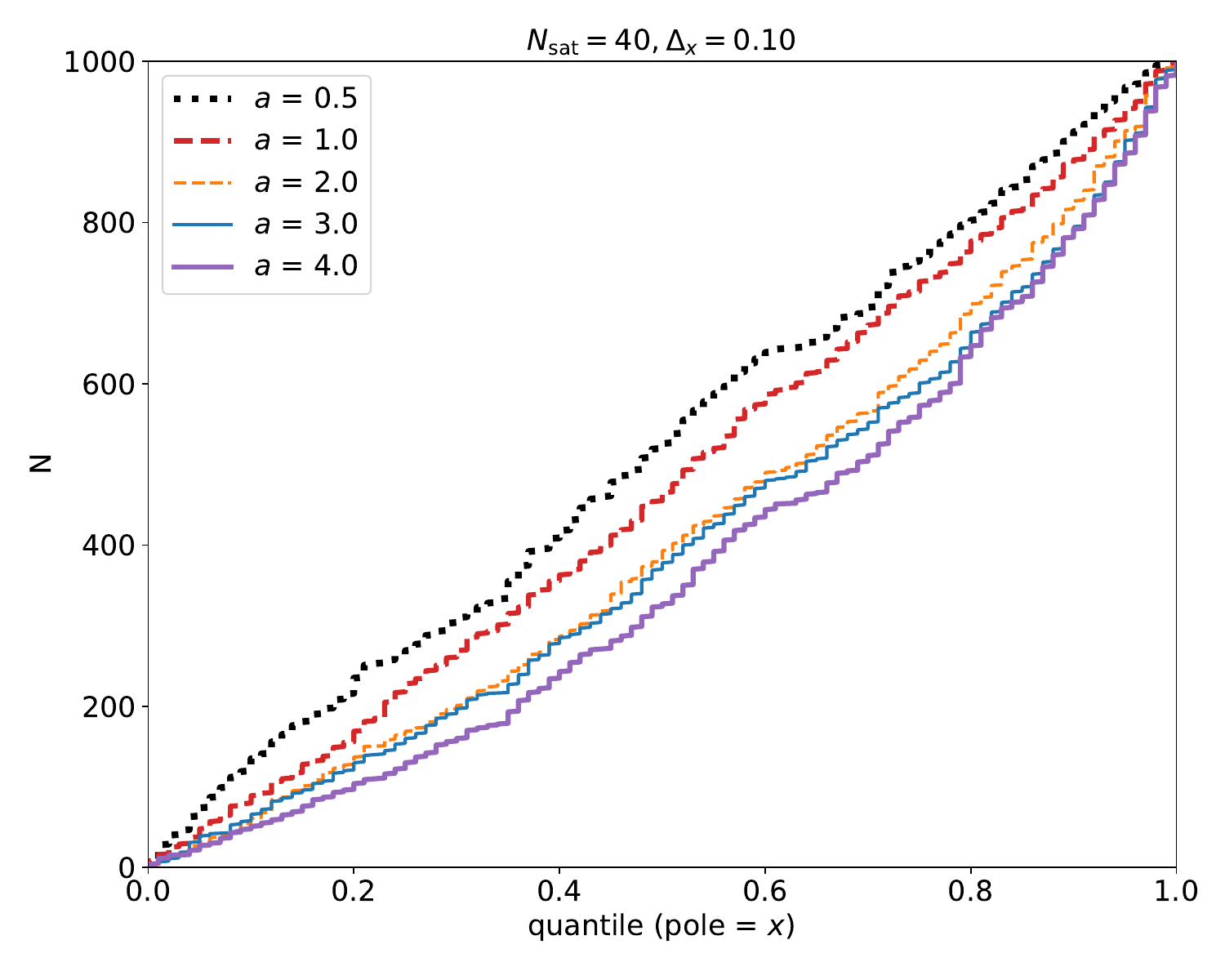}}
      \caption{Effect of radial distribution and lopsidedness on the inferred ''planarity'' of a satellite system using the scaled metric.
      }
         \label{fig:lopsidedness_scaled}
   \end{figure*}

Figure \ref{fig:lopsidedness_scaled} shows that also in case of the scaled metric, the presence of a lopsided satellite distribution biases towards higher inferred degrees of ''planarity'', and that also in this case the effect is stronger for more radially concentrated distributions. Furthermore, contrary to the non-scaled metric, the effect is now present in both considered orientations.

\end{document}